\documentclass[aps,showpacs,amsmath,amssymb]{revtex4}

\usepackage{graphicx}
\usepackage{dcolumn}
\usepackage{bm}
\usepackage{hyperref}

\newcommand{\beq}{\begin{equation}}
\newcommand{\eeq}{\end{equation}}
\newcommand{\beqa}{\begin{eqnarray}}
\newcommand{\eeqa}{\end{eqnarray}}
\newcommand{\beqar}{\begin{eqnarray*}}
\newcommand{\eeqar}{\end{eqnarray*}}
\newcommand{\bra}[1]{\mbox{$\left\langle{#1}\right|$}}
\newcommand{\ket}[1]{\mbox{$\left|{#1}\right\rangle$}}

\def\I{{\rm i}}

\def\e{{\rm e}}

\newcounter{saveeqn}

\begin{document}

\title{Controlled and combined remote implementations \\
of partially unknown quantum operations of multiqubits \\using GHZ
states}
\author{An Min Wang}\email{anmwang@ustc.edu.cn}
\affiliation{Group of Quantum Theory, Department of Modern Physics\\
University of Science and Technology of China, Hefei, 230026,
P.R.China}

\begin{abstract}

We propose and prove protocols of controlled and combined remote
implementations of partially unknown quantum operations belonging to
the restricted sets [An Min Wang: PRA, \textbf{74}, 032317(2006)]
using GHZ states. We detailedly describe the protocols in the cases
of one qubit, respectively, with one controller and with two
senders. Then we extend the protocols to the cases of multiqubits
with many controllers and two senders. Because our protocols have to
demand the controller(s)'s startup and authorization or two senders
together working and cooperations, the controlled and combined
remote implementations of quantum operations definitely can enhance
the security of remote quantum information processing and
potentially have more applications. Moreover, our protocol with two
senders is helpful to farthest arrive at the power of remote
implementations of quantum operations in theory since the different
senders perhaps have different operational resources and different
operational rights in practice.

\end{abstract}

\pacs{03.67.Lx, 03.67.Hk, 03.65.Ud, 03.67.-a}

\maketitle

\section{Introduction}\label{sec1}

Quantum teleportation \cite{Bennett} is one of the most striking
developments in quantum theory. It indicates that a quantum state
can be remotely transferred in a completely different way compared
with a classical state. Thus, one would like to know whether and how
a quantum operation can also be remotely transferred in a completely
different way compared with a classical operation. This problem is
just so-called the remote implementation of quantum operation (RIO),
which was ever studied successfully by Huelga, Plenio and Vaccaro
(HPV) \cite{Huelga1,Huelga2} for the case of one qubit. Recently, we
proposed and proved a protocol of remote implementations of
partially unknown quantum operations of multiqubits via deducing the
general restricted sets and finding the unified recovery operations
\cite{MyRIO}.

Remote implementation of a quantum operation means that this quantum
operation performed on a local system (sneder's) is teleported and
it acts on an unknown state belonging to a remote system
(receiver's) \cite{Huelga1,Huelga2,MyRIO}. Here, a sender is a
person who transfers a quantum operation, and a receiver is a person
whose system receives this quantum operation and this quantum
operation acts on an unknown state belonging to him/her. Obviously,
the RIO is different from simple teleportation of quantum operation
without action, and it is also not an implementation of nonlocal
quantum operation \cite{Cirac,Eisert}, although there are the closed
connections among them. Actually, all of them play the important
roles in distributed quantum computation \cite{Cirac,Eisert},
quantum program \cite{Nielsen,Sorensen} and the other remote quantum
information processing tasks. Recently, a series of works on the
remote implementations of quantum operations appeared and made some
interesting progress both in theory
\cite{Collins,Huelga1,Huelga2,MyRIO} and in experiment
\cite{Huang,Xiang,Huelga3}.

Both HPV's and our recent protocols use Bell states as a quantum
channel. However, it is well-known that GHZ states \cite{GHZ} are
also very important quantum resource in quantum information
processing and communication (QIPC). Just motivated by the scheme of
teleportation of quantum states using GHZ state
\cite{Karlsson,Yang,Deng} and the fact that it has been successfully
applied to quantum secret sharing \cite{Hillery}, we would like to
investigate the remote implementations of quantum operations using
GHZ state(s). Nevertheless, the more important motivations using
state(s)in our protocols are to enhance security, increase variety,
extend applications as well as advance efficiency via fetching in
some controllers and two senders. Our results again indicate that
GHZ states are powerful resources in QIPC.

It is useful and interesting to investigate the remote
implementations of partially unknown quantum operations because they
will consume less overall resource than one of the completely
unknown quantum operations, and such RIOs can satisfy the
requirements of some practical applications. Here, the ``partially
unknown" quantum operations are thought of as those belonging to
some restricted sets which satisfy some given restricted conditions.
In Refs. \cite{Huelga1,Huelga2,MyRIO}, the restricted sets of
quantum operations were seen to be still a very large set of unitary
transformations because their unknown elements take continuous
values. In the simplest case of one qubit, two kinds of restricted
sets of quantum operations are, respectively, a set of diagonal
operations and a set of antidiagonal operations \cite{Huelga2}. For
the cases of multiqubits, we obtained the general forms of
restricted sets of quantum operations and the unified recovery
operations, then proposed and proved our protocol of remote
implementations of quantum operations belonging to the restricted
sets in our recent work \cite{MyRIO}. Specially, our restricted sets
of quantum operations are not reducible to a direct product of two
restricted sets of one-qubit operations, our recovery operations
have general and unified forms, and so our protocol can be thought
of as a development of HPV protocol to multiqubits systems but not a
simple extension of HPV protocol \cite{Huelga2}.

It must be emphasized that the main advantage using GHZ states in
the RIO protocols is to provide ability for fetching in (many)
controller(s) or/and more than one sender. When there is(are) the
controller(s), we call the remote implementation of quantum
operations as controlled one, when there are more than one sender,
we called the remote implementation of quantum operations as
combined one. A controlled remote implementation of quantum
operation has to have the controllers' participation. A combined
remote implementation of quantum operations has to have the senders'
cooperation. Otherwise, the RIOs cannot be faithfully and
determinedly completed.

In the controlled RIOs, not only a controller plays such a role that
the quantum channel between sender and receiver is opened by his/her
operations, but also the controller's measurement (classical
information) impacts the form of the sender's operations or the
receiver's operation. This implies that the controller's action
contains ``start up" and ``authorization" so that the RIOs can be
faithfully and determinedly completed. Just based on this fact, we
can say the controlled RIOs definitely enhance the security of
remote quantum information communication and processing. In
addition, varying with the ways of authorization by the controller,
the steps in RIO protocols have their different forms. It seems to
bring about some complicated expressions of our protocols, but
aiming at the different cases, the controlling process needs such
variety. For example, if the controller trusts in the sender or is
easy to communicate with the sender, he/she authorizes to the
sender; if the controller trusts in the receiver or is easy to
communicate with the receiver he/she authorizes to the receiver; if
the controller hopes to keep the ``say last words", he/she
authorizes to the receiver at a chosen stage of the protocols.

While in the combined RIOs, the later sender has to obtain the
classical information from the former one by one in the sending
sequence of protocols so that he/she can correctly choose his/her
operation. Therefore, the combined RIOs can also definitely enhance
the security of remote quantum information communication and
processing. Note that the security enhancement is in classical
sense, this can be called so-called ``sequential multiple-safety".
This concept can be understood and illustrated by a classical
example of opening safe-deposit box. For simplicity, let us only
consider the case of sequential double-safety. This example is how
to open a safe-deposit box with two locks and its every lock has a
set of various keys. Suppose the set of keys of the first lock are
$k_1^A,k_1^B,\cdots$, the set of keys of the second lock are
$k_2^A,k_2^B,\cdots$. Opening the safe-deposit box needs to use the
sequential and paired keys $(k_1^A,k_2^A)$, or $(k_1^B,k_2^B),
\cdots $ to complete it. Otherwise the safe-deposit box can not be
opened. In other words, two guardians (corresponding two senders)
have to cooperate each other. When the first guardian opens the
first lock using some given key $k_1^C$ (corresponding a quantum
operation belonging to some given restricted set), he/she has to
tell the second guardian his/her using $C$ key so that the second
guardian can correctly use $k_2^C$ to open the safe-deposit box. Of
course, we can say that the combined RIO has higher security. In
addition, in the combined RIOs, our protocol with two senders is
also helpful to farthest arrive at the power of RIOs in theory.
Actually, since it is possible that different senders have different
operational resources and different operational rights in practice,
their cooperations can combine more or more suitable operations and
then our protocol with two senders has a higher practical power of
RIOs than one with only one sender.

Note that in this paper, we only use three partite GHZ states in our
protocols. Therefore, we have at most two senders if only using $N$
GHZ states in the cases of $N$ qubits. In fact, when using more than
three partite GHZ states, we can further extend our protocols to the
cases of more than two senders, even many controllers and many
senders together.

Because the no-cloning/broadcast theorem \cite{Wootters,Barnum}
forbids faithfully to transfer an unknown, even partially (un)known
quantum operation to two locations at the same time, we give up to
consider such a scheme. However, alternatively, we can construct a
symmetric scheme among three parties (locations), in which every
parter plays a role among sender, receiver and controller in the
controlled RIOs, or two parters play two senders and the other
parter plays a receiver in the combined RIOs.

Besides Sec. \ref{sec1} is written as an introduction, this paper is
organized as follows: in Sec.\ref{sec2}, we simply recall the RIO
protocols using Bell states and introduce our restricted sets of
quantum operation of multiqubits; in Sec. \ref{sec3} we propose and
prove protocols of controlled remote implementations of partially
unknown quantum operations of one qubit using one GHZ state; in Sec.
\ref{sec4} we propose and prove a protocol of combined remote
implementations of partially unknown quantum operations of one qubit
using one GHZ state; in Sec. \ref{sec5}, by aid of the explicit form
of our restricted sets of quantum operations of $N$ qubits
\cite{MyRIO} and the general swapping transformations, we extend our
protocols to the cases of multiqubits; in Sec. \ref{sec6}, we
summarize and discuss our results; in appendixes, we analyze the
general swapping transformations used in this paper, and provide the
proofs of our protocols in detail for the cases of more than one
qubits.

\section{RIO using Bell states}\label{sec2}

In the HPV protocol \cite{Huelga1,Huelga2}, the joint system of
Alice and Bob initially reads \beq \label{inis1}\ket{\Psi_{ABY}^{\rm
ini}}=\ket{\Phi^{+}}_{AB}\otimes\ket{\xi}_Y, \eeq where \beq
\ket{\Phi^{+}}_{AB}=\frac{1}{\sqrt{2}}\left(\ket{00}_{AB}+\ket{11}_{AB}\right)
\eeq is one of four Bell states which is shared by Alice (the first
qubit) and Bob (the second qubit), and the unknown state (the third
qubit) \beq \label{1qus} \ket{\xi}_Y=y_0\ket{0}_Y+y_1\ket{1}_Y \eeq
belongs to Bob. Note that the Dirac's vectors with the subscripts
$A,B,Y$ in the above three equations indicate their bases,
respectively, belonging to the qubits $A,B,Y$.

The quantum operation to be remotely implemented belongs to one of
the two restricted sets defined by \beq
U(0,u)=\left(\begin{array}{cc}u_0&
0 \\ 0& u_1\end{array}\right), \quad U(1,u)=\left(\begin{array}{cc} 0 & u_0 \\
u_1 & 0\end{array} \right).\eeq We can say that they are partially
unknown in the sense that the values of their matrix elements are
unknown, but their structures, that is, the positions of their
nonzero matrix elements, are known. In our notation, a restricted
set of one-qubit operations is denoted by $U(d,u)$, where $d=0$ or
$1$ indicates, respectively, this operation belonging to diagonal-
or antidiagonal restricted set, while $u$ is its argument (unknown
elements).

The simplified HPV protocol can be expressed by five steps, which is
made of Bob's preparing, the classical communication from Bob to
Alice, Alice's sending, the classical communication from Alice to
Bob and Bob's recovering \cite{MyRIO}. The whole protocol can be
illustrated by the following quantum circuit: (see Fig.1):
\begin{figure}[ht]
\begin{center}
\includegraphics[scale=0.60]{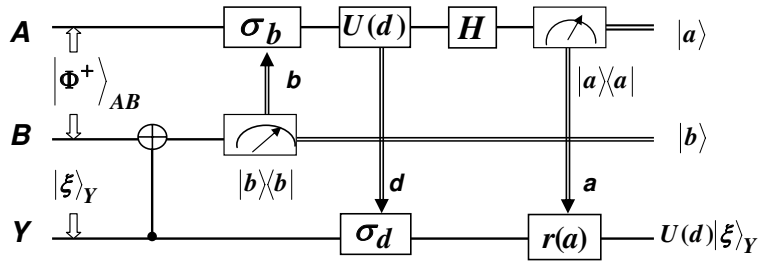}
\end{center}
\vskip -0.1in \caption{Quantum circuit of the simplified HPV
protocol, where $U(d)$ is a quantum operation to be remotely
implemented and it is diagonal or antidiagonal, $H$ is a Hadamard
gate, $\sigma_b,\sigma_d$ are identity matrices or {\sc not} gates
($\sigma_1$) with respect to $b,d=0$ or $b,d=1$, respectively, and
$r(a)=(1-a)\sigma_0+a\sigma_3$ is an identity matrix when $a=0$ or a
phase gate ($\sigma_3$) when $a=1$. The measurements
$\ket{a}\bra{a}$ and $\ket{b}\bra{b}$ are carried out in the
computational basis ($a,b=0,1$). ``$\Rightarrow$" (crewel with an
arrow) indicates the transmission of classical communication to the
location of the arrow direction.} \label{mypic1}
\end{figure}

In order to extend the RIO protocol to the cases of multiqubits, we
first have to seek for the correct restricted sets of quantum
operations of multiqubits that can be remotely implemented in a
faithful and determined way. Actually, we have obtained their
general and explicit forms in our recent works \cite{MyRIO}, that
is, the restricted sets of quantum operations of $N$ qubits have
such forms that every row and every column of operations belonging
to them only has one nonzero element. Thus, it is easy to know that
there are $2^N!$ restricted sets of operations in the $N$-qubit
systems. Denote the $x$th restricted set by $T^r_N(x,t)$, and its
nonzero element in the $m$th row by $t_m$, we have \beq
T^r_N(x,t)=\sum_{m=1}^{2^N}t_m\ket{m,D}\bra{p_m(x),D}.\eeq Here,
$x=1,2,\cdots,2^N$ and \beq
p(x)=(p_1(x),p_2(x),\cdots,p_{2^N}(x))\eeq is an element belonging
to the set of all permutations for the list $\{1,2,\cdots,2^N\}$.
Moreover, when the requirement of the unitary condition is
introduced, $t_m$ will be taken as $\e^{\I\phi_m}$, and $\phi_m$ is
real.

To remotely implement quantum operations belonging to the above
restricted sets, the sender(s) needs a mapping table which provides
one-to-one mapping from $T^r_N(x,t)$ to a classical information $x$
$(x=1,2\cdots,2^N!)$, and the receiver knows a mapping table which
gives out one-to-one mapping from a classical information $x$
$(x=1,2\cdots,2^N!)$ to $R_N(x)$ defined by \beq
\label{fixedu}{R}_N(x)=
T_N^r(x,0)=\sum_{m=1}^{2^N}\ket{m,D}\bra{p_m(z),D}.\eeq Obviously,
it has the same structure as $T_N^r(x,t)$ to be remotely
implemented, and it is an important part in the final recovery
operation.

For simplicity, let us consider the case of two qubits. It is clear
that there are 24 kinds of restricted sets of quantum operations
that can be remotely implemented. In our RIO protocol, we use two
Bell states $|\Phi^{+}\rangle_{A_{1}B_{1}},
|\Phi^{+}\rangle_{A_{2}B_{2}}$ as the quantum channel, where qubits
$A_{1}, A_{2}$ belong to Alice and $B_{1}, B_{2}$ belong to Bob.
Initially, an unknown state $|\xi\rangle_{Y_{1}Y_{2}}$ also belongs
to Bob. Bob first performs two controlled-{\sc not} ($C^{\rm not}$)
transformation by using $Y_{1}, Y_{2}$ as control qubits and $B_{1},
B_{2}$ as target qubits, respectively. Then he measures his qubits
$B_{1}$ and $B_{2}$ in the computational basis
$|b_{1}\rangle_{B_{1}}\langle
b_{1}|\otimes|b_{2}\rangle_{B_{2}}\langle b_{2}|$, where $b_{1},
b_{2}=0, 1$ and sends the results to Alice. After receiving the two
classical bits, Alice first carries out the quantum operations
$\sigma_{b_{1}}^{A_{1}}\otimes\sigma_{b_{2}}^{A_{2}}$ on her two
qubits $A_{1}, A_{2}$. Next Alice acts $T_{2}(x,t)$ on $A_{1}A_{2}$
and executes two Hadamard gate transformation $H_{A_{1}}\otimes
H_{A_{2}}$. Then, she measures her two qubits in the basis
$|a_{1}\rangle_{A_{1}}\langle
a_{1}|\otimes|a_{2}\rangle_{A_{2}}\langle a_{2}|(a_{1}, a_{2}=0, 1)$
and sends the results $a_1a_2$ and $x$ to Bob. As we have mentioned,
the transmission of $x$ is to let Bob choose $R_{2}(x)$ correctly.
With these information, Bob's recovery operations are taken as $
\left[\mathfrak{r}^{Y_1}(a_1)\otimes\mathfrak{r}^{Y_2}(a_2)\right]
{R}_2(x)$, where $\mathfrak{r}(y)=(1-y) \sigma_0+y\sigma_3$.
Finally, our protocol is completed faithfully and determinedly
through the above five steps.

\section{Controlled RIO in the cases of one qubit using one GHZ
state}\label{sec3}

Now, let us first investigate the controlled remote implementations
of quantum operations belonging to restricted sets of one qubit
using one GHZ states. Without loss of generality, we can write the
initial state in a symmetric form of three partite subsystems:
\beq\label{crisi} \ket{\Psi^{\rm ini}}=\ket{{\rm
GHZ}_+}_{ABC}\ket{\chi}_X\ket{\xi}_Y\ket{\zeta}_Z ,\eeq where the
GHZ state has the form \beq \ket{{\rm
GHZ}_+}=\frac{1}{\sqrt{2}}\left(\ket{000}+\ket{111}\right).\eeq It
is shared by Alice, Bob and Charlie. While $\ket{\chi}_X$,
$\ket{\xi}_Y$ and $\ket{\zeta}_Z$ are all unknown states of one
qubit system. The six qubits of the joint system are divided into
three pairs, in which, the qubits $A$ and $X$ belong to Alice, the
qubits $B$ and $Y$ belong to Bob, and the qubits $C$ and $Z$ belong
to Charlie. Obviously, their roles are initially symmetric for the
remote implementations of quantum operations of one qubit.

In order to clearly express our protocol and strictly prove it,
specially, for the cases of multiqubits, we denote that the Hilbert
space of the joint system is initially taken as a direct product of
all qubit Hilbert spaces according to the following sequence: \beq
H=H_A\otimes H_B\otimes H_C\otimes H_X\otimes H_Y\otimes H_Z.\eeq We
can simply call this direct-product ``space structure" and denote it
by a bit-string, for example, the space structure of the above
Hilbert space is $ABCXYZ$. Note that the above space structure is
only a notation rule used here, it is absolutely not a precondition
of the protocols. If we would like to prove our protocols generally
for the cases of multiqubits, such a kind of notation rule is
convenient. This fact can be seen in Appendix A. Obviously, since
taking such a space structure, the subspaces belonging to Alice, or
Bob, or Charlie are separated. This will lead to inconvenience in
the whole-space expression of local operations. Therefore, there is
the need to change the space structure. This can be realized by a
series of swapping transformations, which is studied in Appendix A.

In our protocols, in spite that only local operations and classical
communication are used, the problems we deal with are related with
the whole system because there is entanglement among various partite
subsystems. However, knowing the space structure will be helpful for
us to understand the effect of local operations. In fact, our
protocol can be found due partially to the reasons that we clearly
express an appreciate space structure and general swapping
transformations. Therefore, in the following we keep the above
sequence of direct products of qubit spaces via the whole-space
expressions of our formula in the joint system.

From the symmetric initial state (\ref{crisi}), it is easy to find
that any one partite subsystem of them plays a possible role among
of a sender, a receiver and a controller in the protocols. In other
words, when a controller is fixed to a given partite subsystem,
thus, the other two partite subsystems play a sender and a receiver,
respectively. Under a controller's permission, the remote
implementations of quantum operations belonging to the restricted
sets is faithfully and determinedly completed between the other two
subsystems (locations).

Actually, we are always able to swap their positions in a given
space structure among three partite subsystems using so-called
general swapping transformations that are studied in Appendix A.
Without loss of generality, as soon as a controller is chosen or
dominated, we can rewrite the initial state space structure as \beq
\label{stc1} H_{\rm Controller}\otimes H_{\rm Sender}\otimes H_{\rm
Receiver}\otimes H_{\rm Unknown\;\; State}. \eeq This means that the
first qubit belongs to the controller, the second qubit is in the
sender's partite subsystem (the local subsystem), the third qubit is
mastered by the receiver, and the fourth qubit is an unknown state
in the receiver's hands. Obviously, the unknown states belong to
sender and controller are needless in the protocol as soon as the
roles of attendees are fixed.

When introducing a controller, our protocol of controlled remote
implementations of quantum operations belonging to the restricted
sets is made up of seven steps, in which, there are four steps of
quantum operations including measurement and three times of
classical communications. They are described as the following:

{\em Controlling Step}: This step is carried out by the controller.
He/She performs a Hadamard transformation \beq
H=\frac{1}{\sqrt{2}}\left(\begin{array}{cc}1 &1
\\ 1 & -1\end{array}\right) \eeq on his/her controlled qubit
$\ket{\gamma}$, and then measures it in the computational basis
$\ket{\gamma}\bra{\gamma}\; (\gamma=0,1)$, that is \beqa
\label{cstep} \mathcal{C}(\gamma) &=&
\left(\ket{\gamma}\bra{\gamma}H\right)\otimes I_8, \eeqa where $I_m$
is a $m$ dimensional identity matrix.

This step is a key matter in our protocol. In fact, when the
controller has not done it, there is no quantum entanglement between
any two partite subsystems, so there is no any feasible remote
implementations of operations. Only if a controller agrees or wishes
that the other two partite subsystems implement the RIO protocol,
he/she carries out this operation and measurement. Its action is to
open the quantum channel between the sender and receiver that is
necessary for the remote implementation of operations belonging to
the restricted sets in a faithful and determined way.

{\em Allowing Step}: This step is still completed by the controller,
that is, he/she transfers one ${c}$-bit $\bm{\gamma}$ to the sender
or the receiver, which is denoted by $C_{\rm cs}(\gamma)$ or $C_{\rm
cr}(\gamma)$, respectively.

This allowing step as well as the above controlling step can be
arranged at any time in the RIO process, however, the different
arrangement will result in influences on the steps of our protocols.
If the classical bit $\gamma$ is arranged to transfer to the sender,
this communication has to be done before the other parts of sending
operations. If the classical bit $\gamma$ is chosen to transfer to
the receiver, this communication is able to be done at the beginning
(before receiver's preparation), or in the middle (before the
recovered operations), or at the end (after the standard recovered
operations). At these cases, although the receiver can have the
different choices to finish the protocol, we prefer to use a unified
method for the RIO of one qubit, that is, we use the classical
information $\gamma$ before the end of our protocol.

This step can be understood figuratively as that the controller
distributes the ``password" $\gamma$ to one of the sender and
receiver, or gives an authorization to one of them, or says last
word (to the receiver) in our protocols. This indicates the role of
controller is very important and indispensable. Without the password
distribution by the controller, the sender and receiver cannot
faithfully and determinedly complete the RIO. This can be clearly
seen in the following proofs about our protocols.

{\em Preparing Step}: This step is carried out by the receiver.
There are two kinds of cases, respectively, based on whether the
classical information from the controller is obtained by the
receiver or not.

$\diamond$ Case one: If the receiver does not obtain the classical
information from the controller, he/she first performs a
controlled-{\sc not} using his/her the qubit occupied by the unknown
state as a control, his/her shared part (the qubit $\ket{\beta}$) of
the GHZ state as a target, and then measures his/her shared part of
the GHZ state in the computational basis $\ket{\beta}\bra{\beta}\;
(\beta=0,1)$, that is \beqa \label{rpstep}
\mathcal{P}(\beta)&=&I_4\otimes\left[\left(\ket{\beta}\bra{\beta}\right)
\otimes\sigma_0\right]\left[\sigma_0\otimes
\left(\ket{0}\bra{0}\right)+
\sigma_1\otimes\left(\ket{1}\bra{1}\right)\right]\nonumber \\
&=& I_4\otimes\left[\left(\ket{\beta}\bra{\beta}\right)
\otimes\sigma_0\right]C^{\rm not}(2,1), \eeqa where $\sigma_0$ is
$2\times 2$ identity matrix and $\sigma_i$ $(i=1,2,3)$ are the Pauli
matrices, and $C^{\rm not}$ is a controlled-{\sc not} defined by
\beq\label{dcnot} C^{\rm
not}(2,1)=\sigma_0\otimes\left(\ket{0}\bra{0}\right)
+\sigma_1\otimes\left(\ket{1}\bra{1}\right)=\left(\begin{array}{cccc}1&0&0&0\\
0&0&0&1\\
0&0&1&0\\
0&1&0&0\end{array}\right).\eeq while $(2,1)$, as the variable of
$C^{\rm not}$, indicates that the second qubit is a control and the
first qubit is a target.

The purpose of this step is to let the unknown state be correlated
with the sender's local qubit. This is a precondition that the
sender is able to remotely implement a quantum operation belonging
to the restricted sets.

$\diamond$ Case two: If the controlling step has happened and the
receiver gets the classical bit $\gamma$ from the controller, the
preparing step has three different forms according to the time to
obtain the classical bit $c$ in general.

(1) When the classical bit $\gamma$ is known at the beginning of
this step, the receiver has to add a prior operation \beq
\mathcal{P}^{\rm
pre}(\gamma)=I_4\otimes\left[(1-\gamma)\sigma_0+\gamma\sigma_3\right]
\otimes\sigma_0=I_4\otimes\mathfrak{r}(\gamma)\otimes\sigma_0 \eeq
before the above operation (\ref{rpstep}). Here $\mathfrak{r}(a)$ is
a diagonal phase gate with a parameter that is defined by \beq
\label{rpg}\mathfrak{r}(z)=(1-z)\sigma_0+z\sigma_3=\left(\begin{array}{cc}
1&0\\0&1-2z\end{array}\right)=\left(\begin{array}{cc}
1&0\\0&(-1)^z\end{array}\right).\eeq Note that $z=0,1$, and then
$(-1)^z=1-2z$. Of course, since it commutes with the project
measurement, it also can be inserted between the measurement and the
controlled not in the operation (\ref{rpstep}).

(2) When the classical bit $\gamma$ is known after the operation
(\ref{rpstep}) or before next recovery operation, the receiver
performs a supplementary operation \beq
\label{rpstepaft}\mathcal{P}^{\rm
aft}(\gamma)=I_4\otimes\mathfrak{r}(\gamma)\otimes\mathfrak{r}(\gamma),\eeq
where we have used the fact that \beq
\label{rccnot}\left(\mathfrak{r}(\gamma)\otimes
\mathfrak{r}(\gamma)\right)\left[\left(\ket{\beta}\bra{\beta}\otimes\sigma_0\right)C^{\rm
not}(2,1)\right]=\left[\left(\ket{\beta}\bra{\beta}\otimes\sigma_0\right)C^{\rm
not}(2,1)\right]\left(\mathfrak{r}(\gamma)\otimes\sigma_0\right).\eeq

(3) When the classical bit $\gamma$ is known after the next recovery
operation, this case is discussed putting in the following recovery
step. It is clear that for the above two cases, the receiver always
can delay using the classical information up to after recovery
operation. Therefore, this case is more general. However, it will be
seen that the delaying method is able to lead in the unexpected
complication in the recovery step for the cases of multiqubits.

{\em Classical Communication from receiver to the sender}: This step
is that the receiver transfers a $c$-bit $\bm{\beta}$ to the sender,
which is denoted by $C_{\rm rs}(\beta)$.

The aim of this step is that the receiver tells the sender that
he/she is ready for receiving the remote operation, as well as
his/her preparing way.

It must be emphasized that for the cases of one qubit, the
receiver's preparing has two equivalent ways with respect to
$\beta=0$ or $1$, respectively. If the receiver first fixes the
value of $\beta$ and tells the sender before the beginning of
protocol, this step can be saved. In particular, when $\beta$ is
just taken as $0$, the sender also does not need the transformation
$\sigma_{\beta}$ in the next step, since $\sigma_{0}$ is trivial.

{\em Sending step}: This step is carried out by the sender. There
are two cases.

$\diamond$ Case one: There is no classical information transferred
from the controller to the sender. Thus, after receiving a classical
bit $\beta$ from the receiver, the sender carries out her/his
sending operations which are made of four parts (or five parts in
the case two). The first one is simple $\sigma_\beta$ \beq
\mathcal{P}_S(\beta)=I_2\otimes\sigma_\beta\otimes I_4.\eeq The
second part of sending step is the operation $U(d,u)$ to be remotely
implemented acting on his/her local system (the qubit
$\ket{\alpha}$, a shared part of the GHZ state). The third part of
sending step is simple a Hadamard gate. The fourth, also the final
part is a measurement on the computational basis
$\ket{\alpha}\bra{\alpha}\; (\alpha=0,1)$. All parts of Alice's
sending can be jointly written as \beq \mathcal{S}(\alpha,\beta;
d,u)=\left\{\sigma_0\otimes\left(\ket{\alpha}\bra{\alpha}\right)
\left[H U(d,u)\sigma_\beta\right]\otimes I_4\right\}.\eeq

The action of the first part $\sigma_\beta$ is to perfectly prepare
the state of joint system as such a superposition that the basis in
the locally acted system (belonging to sender's subsystem) of every
component state is the same as its basis in the remotely operated
system (belonging to the space of unknown state in Bob's subsystems)
and the corresponding coefficients are just ones of unknown state.

The second part of sending step is just an operation belonging to
the restricted sets, which will be remotely implemented in our
protocol.

The third part of sending step, the Hadamard gate, is often seen in
quantum computation and quantum communication. Its action is similar
to the cases in the teleportation of states.

The fourth part of sending step is a measurement on the
computational basis whose aim is to project to the needed result.

$\diamond$ Case two: The sender obtains the classical information
$\gamma$ from the controller, he/she has to add to a prior operation
\beq \label{asoadd} \mathcal{S}^{\rm add}(\gamma)=\sigma_0\otimes
\left[(1-\gamma)\sigma_0+\gamma\sigma_3\right]\otimes
I_4=\sigma_0\otimes\mathfrak{r}(\gamma)\otimes I_4\eeq at the
beginning of this step. This means that the sending step becomes
\beq \mathcal{S}^{\rm all}(\alpha,\beta,\gamma;
d,u)=\left\{\sigma_0\otimes\left(\ket{\alpha}\bra{\alpha}\right)
\left[H U(d,u)\sigma_\beta\right]\otimes
I_4\right\}\mathcal{S}_A^{\rm add}(\gamma).\eeq

{\em Classical Communication from sender to receiver}: This step is
that the sender transfers  the classical information $\alpha$ and
$d$ to the receiver, which is denoted by $C_{\rm sr}(\alpha;d)$.

This step is that sender tells the receiver what measurement
(denoted by $\alpha$) has been done and which kind of operations
(denoted by $d$) has been transferred. In our protocol, the sender
has a one to one mapping table to indicate a kind of restrict set by
a value of classical information. For the cases of one qubit, it can
be encoded by one $c$-bit, in which, 0 denotes a restricted set of
diagonal operations and 1 denotes a restricted set of antidiagonal
operations.

{\em Recovery Step}: This step is carried out by the receiver. After
receiving a classical bit $\alpha$ from the sender, the receiver
first requires to do a recovery operation that consists of two parts
or three parts. The first part is $\mathfrak{r}(\alpha)$ and the
second part is a fixed form of a restricted set which has the same
structure as the $U(d,u)$ to be transferred remotely but its nonzero
elements are set as 1. For the cases of one qubit, the fixed forms
of the restricted sets of diagonal- and antidiagonal operations are,
respectively, $\sigma_0$ and $\sigma_1$. Therefore, the receiver's
recovery operations are written as \beq\label{ro}
{\mathcal{R}}(\alpha;d)=I_8\otimes\left[\mathfrak{r}(\alpha)\sigma_d\right],
\eeq where $d=0$ or $1$.

It must be emphasized that the above ${\mathcal{R}}(\alpha;d)$ only
can guarantee the operation $U(d,u)$ is faithfully and determinedly
transferred, if the protocol sets that the controller transfers
his/her classical bit $\gamma$ before its action. Just as statement
above, if $\gamma$ is transferred to the sender, the sender has a
prior preparation part; when $\gamma$ is sent to the receiver before
the his/her preparing step, he/she can add a supplementary part at
the beginning, in the middle or at the end of the preparing step.
Obviously, at the end of the preparing is just before the
recovering, and so we can move this supplementary part to here.
However, if the receiver obtains $\gamma$ from the controller after
the above ${\mathcal{R}}(\alpha;d)$ action, the receiver has to
perform an additional recovery part \beq {\mathcal{R}}^{\rm
aft}(\gamma;d)=(-1)^{\gamma
d}I_4\otimes\mathfrak{r}(\gamma)\otimes\mathfrak{r}(\gamma), \eeq
where $\mathfrak{r}(z)$ is defined in Eq. (\ref{rpg}). Note that $
\mathfrak{r}(\gamma)\ket{\beta}\bra{\beta}=(-1)^{\gamma\beta}\ket{\beta}\bra{\beta}$,
we obtain its another form \beq {\mathcal{R}}^{\rm
aft}(\beta,\gamma;d)=(-1)^{\gamma(
d+\beta)}I_8\otimes\mathfrak{r}(\gamma).\eeq

It is clear that when the protocol sets the controller transfers
his/her classical bit $\gamma$ to the receiver, we always can delay
using the classical information $\gamma$. In other words, in order
to standardize the protocol in the cases of one qubit, we do not add
any $\mathcal{P}^{pre}$ and $\mathcal{P}^{aft}$ in the preparing
step, but we always use the above ${\mathcal{R}}^{\rm aft}$ at the
end of the protocol. Thus, the whole recovery operations are \beq
\mathcal{R}^{\rm all}(\alpha,\gamma,d)={\mathcal{R}}^{\rm
aft}(\gamma;d){\mathcal{R}}(\alpha;d).\eeq However, for the case of
multiqubits, it is not so simple. Generally, we put the additional
recovery operations before the standard recovery operations
(\ref{ro}), even before the preparing step in order to have the
simplest additional operations for the cases of multiqubits. Of
course, if we persist put the additional recovery operation in the
last, we will pay a price that it gets a little complication in
expressing.

In summary, when there is a controller, we have propose four kinds
of protocols for controlled remote implementation of operations
belonging to the restricted sets. (1) the sender obtains password;
(2)-(4) the receiver obtains password respectively before the
preparing, after the preparing (before the recovering) and after the
recovering. The second, third and fourth kinds of our protocols can
be unitedly expressed without obvious difficulty in the case of one
qubit. If the controller transfers his classical bit $\gamma$ to the
sender, the sequence of the above steps in our protocol will become
\beq \mathcal{C}(\gamma)\rightarrow C_{\rm
cs}(\gamma)\rightarrow\mathcal{P}(\beta)\rightarrow C_{\rm
rs}(\beta)\rightarrow\mathcal{S}^{\rm
all}(\alpha,\beta,\gamma;d)\rightarrow C_{\rm
sr}(\alpha;d)\rightarrow\mathcal{R}(\alpha;d), \eeq if the
controller transfers his classical bit $\gamma$ to the receiver, the
sequence of the above steps in our protocol is \beq
\mathcal{C}(\gamma)\rightarrow C_{\rm
cr}(\gamma)\rightarrow\mathcal{P}(\beta)\rightarrow C_{\rm
rs}(\beta)\rightarrow\mathcal{S}(\alpha,\beta;d)\rightarrow C_{\rm
sr}(\alpha;d)\rightarrow\mathcal{R}^{\rm all}(\alpha,\gamma;d). \eeq

It is clear that transferring $\gamma$ to the sender or the receiver
can be figuratively understood as distributing a ``password", in
special, while transferring $\gamma$ to the receiver at the end of
protocol, it can be figuratively understood as ``saying last word".
They are both important controlled means. Besides the password
distributing, the controller owns the right to open the quantum
channel. All of this are some main features of a controlled process.
Therefore, we called the above process as a controlled remote
implementations of operations.

Now, as an example, we fix the controller as Charlie, Alice as a
sender and Bob as a receiver without loss of generality. Thus, the
initial state is simplified as \beq\label{br} \ket{\Psi_{ABCY}^{\rm
ini}}=F^{-1}_4(1,3)\left[\ket{{\rm GHZ}_+}_{CAB}\ket{\xi}_Y\right],
\eeq where $F_3(1,3)$ is a forward rearrangement made of two
swapping transformations between the neighbor qubits, which is
defined in Appendix A. Similarly, we can discuss the other choices
of the controller, sender and receiver, but we do not intend to
discuss here.

If we set that the Charlie (controller) transfers the password to
sender. It is the first kind of our protocols. All of the operations
and measurements in our protocol can be jointedly written as \beqa
\mathcal{I}_R(a,b,c;d)&=&F^{-1}_4(1,3)\left\{\left(\ket{c}_{C}\bra{c}H^C\right)\otimes
\left(\ket{a}_{A}\bra{a}H^AU(d,u)\sigma_b^A\mathfrak{r}(c)\right)\right.\nonumber\\
& &\left.\otimes\left[\left(\sigma_0\otimes
\mathfrak{r}(a)\right)C^{\rm not}\right]\right\}F_4(1,3). \eeqa Note
that the operations with the superscripts $A,C$ denote their Hilbert
spaces belonging, respectively, to the spaces of qubits $A,C$.
Sometimes, if there is no any confusion, we omit these superscripts.
This whole space form of our protocol has covered up the sequence of
operations and steps of classical communication, but its advantage
is clear. Its action on the initial state (\ref{br}) yields \beq
\label{rioghzbr}\ket{\Psi_{ABCY}^{\rm
final}(a,b,c;d)}={\mathcal{I}}_R(a,b,c;d)\ket{\Psi_{ABCY}^{\rm
ini}}= \frac{1}{2\sqrt{2}}\ket{a bc}_{ABC}\otimes U(d,u)\ket{\xi}_Y.
\eeq The unknown state to be remotely implemented is just
$\ket{\xi}_Y$ in Bob's partite subsystem defined by Eq.(\ref{1qus}).
Our protocol is then determinedly and faithfully completed.

When setting that Charlie's information transfers to Alice, the
whole process of controlled remote implementation of quantum
operations belonging to the restricted sets is shown in Fig.2.

\begin{figure}[ht]
\begin{center}
\includegraphics[scale=0.60]{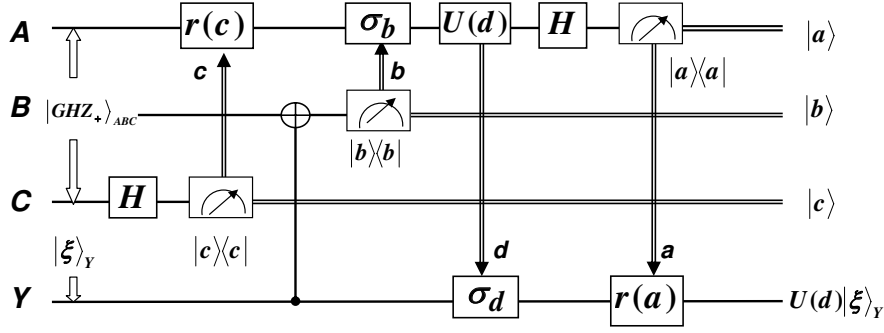}
\end{center}
\vskip -0.3in \caption{Quantum circuit of the controlled remote
implementations of quantum operations with a controller Charlie.
Here, $U(d)$ belonging to the restricted sets is a quantum operation
to be remotely implemented, $H$ is a Hadamard transformation,
$\sigma_b,\sigma_d$ are identity matrices or {\sc not} gates
($\sigma_1$) for $b,d=0$ or $b,d=1$ respectively, and
$r(x)=(1-x)\sigma_0+x \sigma_3$ is equal an identity matrix if $x=0$
or a phase gate if $x=1$. The measurements
$\ket{a}\bra{a}$,$\ket{b}\bra{b}$ and $\ket{c}\bra{c}$ are carried
out in the computational basis ($a,b,c=0,1$). ``$\Rightarrow$"
(crewel with an arrow) indicates the transmission of classical
communication to the location of arrow direction.} \label{mypic1}
\vskip -0.1in
\end{figure}

Here, we only express the full operations for the first kind of our
protocols, and provide its figure of quantum circuit. For the other
three kinds of our protocols, the full operations and the figures of
quantum circuits are similar. In addition, we should notice that the
controller cannot choose who is a sender and who is a receiver in
the other two partite subsystems. In other words, when Charlie is a
controller, either of Alice and Bob can be chosen as a sender and
the other one partite subsystem plays a receiver.

In the end of this section, let us prove our above protocol in
detail. For simplicity, we only consider the cases that Alice is a
sender, Bob is taken as a receiver, and Charlie is a controller.
Initially, the joint system is in the state (\ref{br}). When Charlie
agrees or wishes that Alice and Bob can carry out the remote
implementations of quantum operations belonging the restricted sets,
he will open the quantum channel between them by preforming the
controlling operation on his qubit. His action gives \beqa
\ket{\Psi_{ABCY}^C(c)}&=&F^{-1}_4(1,3)\mathcal{C}(c)F_4(1,3)\ket{\Psi^{\rm
ini}_{ABCY}}\nonumber\\
&=&\frac{1}{2}F^{-1}_4(1,3)
\left[\ket{c}_C\otimes\left(\ket{00}_{AB}+(-1)^c\ket{11}_{AB}\right)
\otimes\ket{\xi}_Y\right]\\
&=&\frac{1}{2}F^{-1}_4(1,3)\left(\sigma_0\otimes\mathfrak{r}(c)\otimes
I_4\right)
\left[\ket{c}_C\otimes\left(\ket{00}_{AB}+\ket{11}_{AB}\right)
\otimes\ket{\xi}_Y\right]\\
&=&\frac{1}{2}F^{-1}_4(1,3)\left(I_4\otimes\mathfrak{r}(c)\otimes
\sigma_0\right)
\left[\ket{c}_C\otimes\left(\ket{00}_{AB}+\ket{11}_{AB}\right)
\otimes\ket{\xi}_Y\right],\eeqa where we have used the definition of
$\mathfrak{r}(y)$ in Eq. (\ref{rpg}). Thus, Alice and Bob now share
a Bell state, and they can carry out the protocol of RIO. However,
because HPV protocol is dependent on the type of Bell state, Charlie
has to send the ``password" $c$ to Alice or Bob. Actually, this
indicates the Charlie has his control right.

We need to consider two cases.

The first case is that the protocol sets Charlie to transfer his
classical information $c$ to Alice. Using $\mathcal{P}(b)$, Bob
prepares his state as \beqa
\ket{\Psi_{ABCY}^P(b,c)}&=&F^{-1}_4(1,3)\mathcal{P}(b)F_4(1,3)\ket{\Psi^{C}_{ABCY}}\nonumber\\
&=&\frac{1}{2}\left[F^{-1}_4(1,3)\left(I_2\otimes\mathfrak{r}(c)\otimes
I_4\right)\right]
\ket{c}_C\otimes\left\{\sigma_0\otimes\left[\left(\ket{b}\bra{b}\otimes\sigma_0\right)C^{\rm
not}\right]\right\}\nonumber\\
& &\times\left[\sum_{k=0}^1
y_k\left(\ket{00k}_{ABY}+\ket{11k}_{ABY}\right)\right].\eeqa Note
that \beqa \label{pbell} & &\left\{\sigma_0\otimes
\left[\left(\ket{b}\bra{b}\right) C^{\rm
not}(2,1)\right]\right\}\left[
\left(\ket{00k}+\ket{11k}\right)\right]\nonumber\\
&=&\left[\sigma_0\otimes \left(\ket{b}\bra{b}\right)
\otimes\sigma_0\right]
\left\{\left(\ket{000}+\ket{110}\right)\delta_{k0}
+\left(\ket{011}+\ket{101}\right)\delta_{k1}\right\}\nonumber\\
&=&\left(\ket{0b0}\delta_{b0}+\ket{1b0}\delta_{b1}\right)\delta_{k0}
+\left(\ket{0b1}\delta_{b1}+\ket{1b1}\delta_{b0}\right)\delta_{k1}\nonumber\\
&=& \left[ \ket{bb0}\left(\delta_{b0}+\delta_{b1}\right)\delta_{k0}
+\ket{(1-b)b1}\left(\delta_{b1}+\delta_{b0}\right)\delta_{k1}\right]\nonumber\\
&=&\left(\sigma_{b}\otimes I_4\right)\left(\delta_{k0}\ket{0b0}
+\delta_{k1}\ket{1b1}\right)
\nonumber\\
&=&\left(\sigma_{b}\otimes
I_4\right)\left(\delta_{k0}+\delta_{k1}\right)\ket{kbk}\nonumber\\
&=& \left(\sigma_{b}\otimes I_4\right)\ket{kbk}, \eeqa where we have
used the facts that $\sigma_b\ket{b}=\ket{0}$ and
$\sigma_b\ket{1-b}=\ket{1}$ for $b=0,1$. It results in \beqa
\ket{\Psi_{ABCY}^P(b,c)}&=&
\frac{1}{2}\left[F^{-1}_4(2,4)\left(\mathfrak{r}(c)\sigma_b\otimes
I_8\right)\right]\left[\left(y_0\ket{00}_{AY}+y_1\ket{11}_{AY}\right)\otimes
\ket{bc}_{BC}\right].\eeqa

After Bob is ready, he transfers a classical bit $b$ in order to
tell Alice his preparing way. So Alice starts with a supplementary
operation so that the state of joint system is perfectly ready via
the (\ref{asoadd}), that is \beqa
\ket{\Psi^P_f(b,c)}&=&F^{-1}_4(1,3)\mathcal{S}_A^{\rm
add}(c)F_4(1,3)\ket{\Psi^{P}_{ABCY}}\nonumber\\&=&\frac{1}{2}F^{-1}_4(2,4)
\left[\left(\sigma_b\otimes\sigma_0\right)\left(y_{0}
\ket{00}_{AY}+y_{1}\ket{11}_{AY}\right) \otimes \ket{b}_B\ket{c}_C
\right].\eeqa Therefore, Alice's sending step yields \beqa
\label{aftersending}\ket{\Psi^S(a,b,
c;d)}&=&F^{-1}_4(1,3)\mathcal{S}^{\rm
all}(a,b,c,d)F_4(1,3)\ket{\Psi^{P}_{ABCY}}\nonumber\\
&=&\frac{1}{2}F^{-1}_4(2,4)\left\{\left[\sum_{k}^1y_{k}\bra{a}HU(d,u)\ket{k}
\ket{a}_{A}\ket{k}_Y\right]\otimes \ket{bc}_{BC}\right\}.\eeqa

The second case is that the protocol sets Charlie to transfer his
classical information $c$ to Bob. If Bob choose to first perform
$\mathcal{P}^{\rm pre}$, then its action is the same as Alice's.
Therefore, when Alice finishes the sending operation, we also obtain
Eq. (\ref{aftersending}). Note that $\mathcal{P}^{\rm
aft}(c)\mathcal{P}(b)=\mathcal{P}(b)\mathcal{P}^{\rm pre}(c)$, we
can, after $\mathcal{P}(b)$ acting, use $\mathcal{P}^{\rm aft}(c)$.
From $\mathcal{R}(a;d)\mathcal{P}^{\rm
aft}(c)=(-1)^{cd}\mathcal{P}^{\rm aft}(c)\mathcal{R}(a;d)$, this
also means that Bob can delay the additional recovery operation to
the end. It is clear that the results of three kinds of procedures
are the same.

Now, Bob performs recovery operation (\ref{ro}). From the relation
that $U(d,u)=U(0,u)\sigma_d $, and the facts that
$U(0,u)=\sum_{j=0}^1u_j\ket{j}\bra{j}$ and
$\mathfrak{r}(a)=\sum_{l=0}^1 (-1)^{a l}\ket{l}\bra{l}$, it follows
that \beqa \label{}\ket{\Psi^{\rm final}(a,b,c;d)}&=&F^{-1}_4(1,3)
\mathcal{R}(a,d)F_4(1,3)\ket{\Psi^{S}_{ABCY}}\nonumber\\
&=&\frac{1}{2}F^{-1}_4(2,4)\left\{\left[\sum_{k}^1y_{k}\bra{a}HU(d,u)\ket{k}
\ket{a}_{A}\left(\mathfrak{r}(a)\sigma_d\ket{k}_Y\right)\right]
\otimes \ket{bc}_{BC}\right\}\nonumber\\
&=& \frac{1}{2}\ket{a}_A\otimes
F^{-1}_3(1,3)\left\{\left[\sum_{j=0}^1\sum_{k=0}^1\sum_{l=0}^1
u_{j}y_{k}\bra{a}H\ket{j}\right.\right.\nonumber\\
& &\left.\left.\times\left(\bra{j}\sigma_d\ket{k}(-1)^{a l}\bra{l}
\sigma_d\ket{k}\right)\ket{l}_{Y}\right]\ket{bc}_{BC}\right\}.\eeqa
Note that \beq \label{riopr1}\bra{j}\sigma_d\ket{k}
\bra{l}\sigma_d\ket{k}=\bra{j}\sigma_d\ket{k}\delta_{jl},\eeq we
have \beqa \ket{\Psi^{\rm final}(a,b,c;d)}&=&
\frac{1}{2}\ket{a}_A\otimes
F^{-1}_3(1,3)\left\{\left[\sum_{j=0}^1\sum_{k=0}^1 u_{j}y_{k}(-1)^{a
j}\bra{a}H\ket{j}\bra{j}\sigma_d\ket{k} \ket{j}_Y\right]\otimes
\ket{bc}_{BC}\right\}.\eeqa Since \beq \label{riopr2}(-1)^{a
j}\bra{a}H\ket{j} =\frac{1}{\sqrt{2}}\eeq for any $a$ and $j$, the
above equation becomes \beqa \ket{\Psi^{\rm final}(a,b,c;d)}&=&
\frac{1}{2\sqrt{2}}\ket{a}_A\otimes
F^{-1}_3(1,3)\left\{\left[\sum_{j=0}^1\sum_{k=0}^1
u_{j}y_{k}\bra{j}\sigma_d\ket{k} \ket{j}_Y\right]\otimes
\ket{bc}_{BC}\right\}\nonumber\\
&=&\frac{1}{2\sqrt{2}}\ket{a}_A\otimes
F^{-1}_3(1,3)\left\{\left[\sum_{k=0}^1
y_{k}U(0,u)\sigma_d\ket{k}_Y\right]\otimes
\ket{bc}_{BC}\right\}\nonumber\\
&=&\frac{1}{2\sqrt{2}}\ket{abc}_{ABC} \otimes U(d,u)\ket{\xi}_Y.
\eeqa That is, we obtain the conclusion (\ref{rioghzbr}) of our
protocol. Therefore, we finish the proof our protocols of controlled
RIO with a controller in the cases of one qubit.

\section{Combined RIO in the case of one qubit using one GHZ
state}\label{sec4}

Now, let us consider a quantum operation that is a product of two
parts $\mathcal{U}_2$ and $\mathcal{U}_1$, that is,
$\mathcal{U}=\mathcal{U}_2 \mathcal{U}_1$. Assuming $\mathcal{U}_1$
and $\mathcal{U}_2$ both belong to the restricted sets,  we can
denote them by $U(d_1,u)$ and $U(d_2,v)$, respectively, in our
notation. Thus, the remote implementation of $\mathcal{U}$ can be
completed via sending $\mathcal{U}_1$ and $\mathcal{U}_2$ in turn by
one sender in the known protocols \cite{Huelga2,MyRIO}, but two
shared Bell pairs are needed. However, we find that this task can be
faithfully and determinedly completed by two senders via one GHZ
state. Moreover, we will see that the RIO protocol with two senders
using one GHZ state has higher security compared with one using two
Bell states. More analysis about the security enhancement has been
given in our introduction.

Without loss of generality, we set Alice and Bob as two senders, and
Alice first sends $U(d_1,u)$, Bob then sends $U(d_2,v)$. Charlie
plays a receiver. Except for the unknown state is replaced  by
$\ket{\zeta}=z_0\ket{0}_Z+z_1\ket{1}_Z$, the initial state has not
the other difference form Eq. (\ref{br}). Since the significance and
actions of the most related operations have been explained in Sec.
\ref{sec3}, we do not intend to repeat them here. Our protocol is
made of the following seven steps.

{\em Step one: Charlie's preparing}. \beq
\mathcal{P}_C(c)=I_4\otimes\left[\left(\ket{c}_C\bra{c}\right)\otimes\sigma_0^Z\right]
\left[\sigma_0^C\otimes \left(\ket{0}_Z\bra{0}\right)+
\sigma_1^C\otimes\left(\ket{1}_Z\bra{1}\right)\right]. \eeq

{\em Step two: First classical communication}. Charlie sends the
classical information $c$ to Alice and Bob.

{\em Step three: Alice's sending}. \beq \mathcal{S}_A(a,c;
d_1,u)=\left(\ket{a}_A\bra{a}\right)\left[H^A
U(d_1,u)\right]\sigma_c^A\otimes I_8.\eeq

{\em Step four: Second classical communication}. Alice sends the
classical information $d_1$ to Bob and $a$ to Charlie.

{\em Step five: Bob's sending}. \beq
\mathcal{S}_B(b,c;d_1,d_2,v)=\sigma_0\otimes\left(\ket{b}_B\bra{b}\right)
\left[H^B U(d_2,v)\right]\left(\sigma_{d_1}^B \sigma_c^B
\right)\otimes I_4.\eeq

{\em Step six: Third classical communication}. Bob sends the
classical information $b$ and $d_2$ to Charlie.

{\em Step seven: Charlie's recovering} \beq
{\mathcal{R}}_C(a,b;d_1,d_2)=I_4\otimes\sigma_0^C\otimes
\left(\mathfrak{r}(b)\sigma_{d_2}\right).\left(\mathfrak{r}(a)\sigma_{d_1}\right).\eeq
All of the operations and measurements in our above protocol can be
jointly written as \beqa \mathcal{I}_R(a,b,c;d_1,d_2,u,v)&=&
\left(\ket{a}_{A}\bra{a}H^AU(d_1,u)\sigma_c^A\right)\otimes
\left(\ket{b}_{B}\bra{b}H^BU(d_2,v)\sigma_{d_1}\sigma_c^A\right)
\nonumber\\
& &\times\left[\left(\ket{c}\bra{c}\otimes
\mathfrak{r}(b)\sigma_{d_2}\mathfrak{r}(a)\sigma_{d_1}\right)C^{\rm
not}\right]. \eeqa

Its acting on the initial state gives \beq \ket{\Psi_{C}^{\rm
final}(a,b,c;d_1,d_2,u,v)}={\mathcal{I}}_R(a,b,c;d_1,d_2,u,v)\ket{\Psi_{C}^{\rm
ini}}= \frac{1}{2\sqrt{2}}\ket{a b c}_{ABC}\otimes
U(d_2,v)U(d_1,u)\ket{\zeta}_Z, \eeq where $a,b,c=0,1$ denote the
spin up or down, and $d_1, d_2=0$ and $d_1,d_2=1$ respectively
indicate the operations of diagonal and antidiagonal restricted
sets. Therefore, the remote implementations of the combination of
two quantum operations belonging to restricted sets are faithfully
and determinedly completed. It can be called the combined remote
implementation of quantum operations which can be displayed by
Fig.3.

\begin{figure}[ht]
\begin{center}
\includegraphics[scale=0.60]{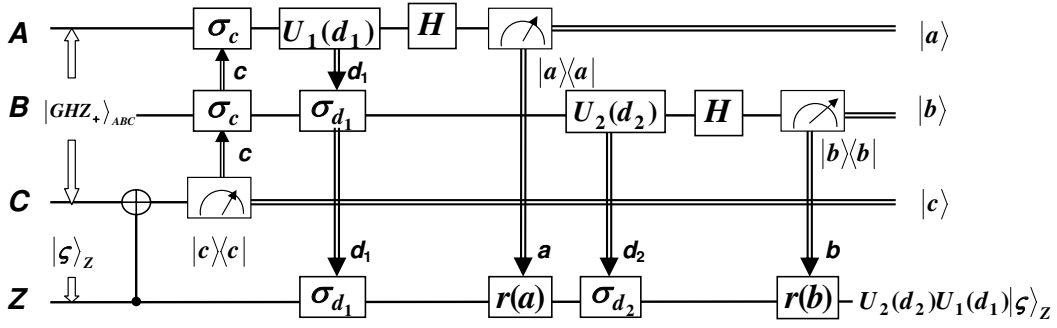}
\end{center}
\vskip -0.3in \caption{Quantum circuit of the combined remote
implementation of quantum operation with two sender Alice and Bob .
Here, $U_1(d_1)$ and $U_2(d_2)$ are respectively a part of the
quantum operation $U(d_1,d_2)=U_2(d_2) U_1(d_1)$ that is remotely
implemented by combining Alice and Bob's actions, $H$ is a Hadamard
transformation, $\sigma_c,\sigma_{d_1},\sigma_{d_2}$ are identity
matrices or {\sc not} gates ($\sigma_1$) for $c,d_1,d_2=0$ or
$c,d_1,d_2=1$ respectively, and $r(x)=(1-x)\sigma_0+x \sigma_3$ is
an identity matrix if $x=0$ or a phase gate ($\sigma_3$) if $x=1$.
The measurements $\ket{a}\bra{a}$,$\ket{b}\bra{b}$ and
$\ket{c}\bra{c}$ are carried out in the computational basis
($a,b,c=0,1$). ``$\Rightarrow$" (crewel with an arrow) indicates the
transmission of classical communication to the location of arrow
direction.} \label{mypic2}\vskip -0.1in
\end{figure}

In order to prove our protocol with two senders, we first need the
equation \beq \label{pghz}\left\{\sigma_0\otimes
\left[\left(\ket{c}\bra{c}\otimes\sigma_0\right) C^{\rm
not}(2,1)\right]\right\}\ket{\rm GHZ_+}\otimes\ket{k}
=F^{-1}_4(3,4)\left(\sigma_c\otimes\sigma_c\otimes
I_4\right)\left(\ket{kkk}\otimes\ket{c}\right). \eeq Its proof is
similar to Eq. (\ref{pbell}). Therefore, Charlie's preparation gives
\beqa \ket{\Psi^P(c)}&=&\mathcal{P}_C(c)\ket{\Psi^{\rm
ini}_{ABCZ}}\nonumber\\
&=&\frac{1}{\sqrt{2}}F^{-1}_4(3,4)\left[\left(\sigma_c\otimes\sigma_c\otimes
I_4\right)\left(z_{0} \ket{00}_{AB} \otimes\ket{0}_{Y}
+z_{1}\ket{11}_{AB} \otimes\ket{1}_{Y}\right) \otimes
\ket{c}_C\right], \eeqa where $F_N(i,j)$ is defined in Appendix A.

After receiving the classical information $c$ from Charlie
(receiver), Alice supplements a $\sigma_c$ transformation, and then
performs the first sending operations, we have \beqa
\ket{\Psi_1^S(a,
c;d_1,u)}&=&\mathcal{S}_A(a,c,d_1,u)\ket{\Psi^P(c)}\nonumber\\
&=&\frac{1}{\sqrt{2}}F^{-1}_4(3,4)
\left[\sum_{k=0}^1z_{k}(\ket{a}_A\bra{a}HU(d_1,u)
\ket{k})\left(\sigma_c\ket{k}_B\right)\ket{k}_{Y}\right]\otimes
\ket{c}_{C}.\eeqa Alice again tells Bob $d_1$ and Charlie $a$. In
succession, based on the classical information $c$ (coming from
Charlie) and $d_1$ (coming from Alice), Bob first carries out
$\sigma_{d_1}\sigma_c$, and then performs the second sending
operations: \beqa \ket{\Psi_2^S(a,b,
c;d_1,d_2,u,v)}&=&\mathcal{S}_B(b,c,d_2,v)\ket{\Psi^S_1(a,c,d_1,u}\nonumber\\
&=&\frac{1}{\sqrt{2}}F^{-1}_4(3,4)
\left[\sum_{k=0}^1z_{k}\left(\ket{a}_A\bra{a}HU(d_1,u)
\ket{k}_{A}\right)\right.\nonumber\\
& &\otimes\left.\left(\ket{b}_B\bra{b}HU(d_2,v)\sigma_{d_1}
\ket{k}_B\right)\ket{k}_{Y}\right]\otimes \ket{c}_{C}.\eeqa Finally,
Bob's recovery operation gives \beqa \ket{\Psi^{\rm final}(a,b,
c;d_1,d_2,u,v)}&=&\mathcal{R}_B(a,b,d_1,d_2)\ket{\Psi^S_2(a,b,c,d_1,d_2,u,v}\nonumber\\
&=& \frac{1}{\sqrt{2}}F^{-1}_4(3,4)
\left[\sum_{k=0}^1z_{k}\left(\ket{a}_A\bra{a}HU(d_1,u)
\ket{k}\right)\right.\nonumber\\
& &\left.\times\left(\ket{b}_B\bra{b}HU(d_2,v)\sigma_{d_1}
\ket{k}\right)\right.\nonumber\\
& &\otimes\left. \mathfrak{r}(b)\sigma_{d_2}
\mathfrak{r}(a)\sigma_{d_1}\ket{k}_{Z}\right]\otimes
\ket{c}_{C}.\eeqa  The remaining steps of this proof is similar to
the ones in the end of Sec. \ref{sec3}, but it needs to repeat three
times. First, using
$U(d_1,u)=U(0,u)\sigma_{d_1}=\sum_{j=0}^1u_j\ket{j}\bra{j}\sigma_{d_1}$
and $\mathfrak{r}(a)=\sum_{l=0}^1 (-1)^{a l}\ket{l}\bra{l}$ we have
\beqa \ket{\Psi^{\rm final}(a,b,
c;d_1,d_2,u,v)}&=&\frac{1}{\sqrt{2}}
\left[\sum_{j=0}^1\sum_{k=0}^1\sum_{l=0}^1u_jy_{k}\left(\ket{a}_A\bra{a}H
\ket{j}\bra{j}\sigma_{d_1}\ket{k}\right)\right.\nonumber\\
& &\otimes\left(\ket{b}_B\bra{b}HU(d_2,v)\sigma_{d_1}
\ket{k} \right)\nonumber\\
& &\left. \otimes\mathfrak{r}(b)\sigma_{d_2}(-1)^{al}
\ket{l}_Y\bra{l}\sigma_{d_1}\ket{k}\right]\otimes \ket{c}_{C}.\eeqa
Because that
$\bra{j}\sigma_{y}\ket{k}\bra{l}\sigma_{d_1}\ket{k}=\delta_{jl}\bra{j}\sigma_{y}\ket{k}$,
it becomes \beqa \ket{\Psi^{\rm final}(a,b,
c;d_1,d_2,u,v)}&=&\frac{1}{\sqrt{2}}\ket{ab}_{AB}\otimes
\left[\sum_{j=0}^1\sum_{k=0}^1u_jy_{k}\left(\bra{a}H
\ket{j}(-1)^{aj}\right)\right.\nonumber\\
& &\times\left(\bra{b}HU(d_2,v)\sigma_{d_1} \ket{k} \right)\left.
\mathfrak{r}(b)\sigma_{d_2}
\ket{j}_Y\bra{j}\sigma_{d_1}\ket{k}\right]\otimes \ket{c}_{C}.\eeqa
While from $\bra{a}H \ket{j}(-1)^{aj}=1/\sqrt{2}$, it follows that
\beqa \ket{\Psi^{\rm final}(a,b,
c;d_1,d_2,u,v)}&=&\frac{1}{2}\ket{ab}_{AB}\otimes
\left[\sum_{j=0}^1\sum_{k=0}^1u_jy_{k}\left(\bra{b}HU(d_2,v)\sigma_{d_1} \ket{k} \right)\right.\nonumber\\
& &\times \left. \mathfrak{r}(b)\sigma_{d_2}
\ket{j}_Y\bra{j}\sigma_{d_1}\ket{k}\right]\otimes \ket{c}_{C}. \eeqa
Again inserted the complete relation after $U(d_2)$ and based on the
above same reason, the above equation is reduced to \beqa
\ket{\Psi^{\rm final}(a,b,
c;d_1,d_2,u,v)}&=&\frac{1}{2}\ket{ab}_{AB}\otimes
\left[\sum_{j=0}^1\sum_{k=0}^1\sum_{l=0}^1u_jy_{k}
\left(\bra{b}HU(d_2,v)\ket{l}\bra{l}\sigma_{d_1} \ket{k} \right)\right.\nonumber\\
& &\left. \times \mathfrak{r}(b)\sigma_{d_2}
\ket{j}_Y\bra{j}\sigma_{d_1}\ket{k}\right]\otimes \ket{c}_{C}
\nonumber\\
&=&\frac{1}{2}\ket{ab}_{AB}
\otimes\left[\sum_{j=0}^1\sum_{k=0}^1u_jy_{k}
\left(\bra{b}HU(d_2,v)\ket{j}\right)\right.\nonumber\\
& &\left. \times\mathfrak{r}(b)\sigma_{d_2}
\ket{j}_Y\bra{j}\sigma_{d_1}\ket{k}\right]\otimes \ket{c}_{C}. \eeqa
Repeatedly using the above skills, that is
$U(d_2)=U(0,v)\sigma_{d_2}=\sum_{i=0}^1v_i\ket{i}\bra{i}\sigma_{d_2}$
and $\mathfrak{r}(b)=\sum_{l=0}^1(-1)^{bl}\ket{l}\bra{l}$, we obtain
\beqa \ket{\Psi^{\rm final}(a,b, c;d_1,d_2,u,v)}
&=&\frac{1}{2}\ket{ab}_{AB}\otimes
\left[\sum_{i=0}^1\sum_{j=0}^1\sum_{k=0}^1\sum_{l=0}^1v_iu_jy_{k}
\left(\bra{b}H\ket{i}\bra{i}\sigma_{d_2}\ket{j}\right)\right.\nonumber\\
& &\left. \times (-1)^{bl}\ket{l}_Y\bra{l}\sigma_{d_2}
\ket{j}\bra{j}\sigma_{d_1}\ket{k}\right]\otimes \ket{c}_{C}\nonumber\\
&=&\frac{1}{2}\ket{ab}_{AB}\otimes
\left[\sum_{i=0}^1\sum_{j=0}^1\sum_{k=0}^1v_iu_jy_{k}
\left(\bra{b}H\ket{i}(-1)^{b i}\right)\right.\nonumber\\
& &\left. \times\bra{i}\sigma_{d_2}
\ket{j}\bra{j}\sigma_{d_1}\ket{k}\ket{i}_Y\right]\otimes \ket{c}_{C}
\nonumber\\
&=&\frac{1}{2\sqrt{2}}\ket{ab}_{AB}\otimes
\left[\sum_{i=0}^1\sum_{j=0}^1\sum_{k=0}^1v_iu_jy_{k}
\bra{i}\sigma_{d_2}
\ket{j}\bra{j}\sigma_{d_1}\ket{k}\ket{i}_Y\right]\otimes
\ket{c}_{C}\nonumber\\
&=&\frac{1}{2\sqrt{2}}\ket{ab}_{AB}\otimes
\left[U(d_2,v)U(d_1,u)\ket{\xi}_Y\right]\otimes \ket{c}_{C}.\eeqa
The proof of our protocol with two senders using one GHZ state is
finished.

\section{Protocols in the cases of $N$ qubits}\label{sec5}

We have proposed and proved the protocols of controlled and combined
remote implementations of quantum operations for one qubit using one
GHZ state, respectively. The controlled RIO has a controller, a
sender and a receiver, and the roles of three partite subsystems can
be exchanged with each other. The combined RIO has two senders and a
receiver, and every sender transfers an operation (or a part of one
decomposable operation), two transferred operations are combined
together according to the sequence of transferred time to form a
total operation that is remotely implemented. In the following let
us investigate the cases of multiqubits.

We have seen that since a controller or an extra sender being
fetched in the protocols for the cases of one qubit, the
entanglement resource really used to the remote implementations of
operations is still one Bell pair, in spite that our protocols is
carried out via one GHZ state. In other words, the design of adding
controller or extra sender will use a part of the entanglement
resource. Therefore, for the cases of more than one qubit, our
restricted sets \cite{MyRIO} are still suitable to our controlled
and combined RIOs.

\subsection{Some notations}

Usually, in order to avoid the possible errors, we need to denote
the sequential structure of direct product space of qubits, or a
sequence of direct products of qubit space basis vectors in the
multiqubit systems. For Alice's space, we set its sequential
structure as $A_1A_2\cdots A_N$, in other words, its basis vector
has the form $\ket{a_1}_{A_1}\ket{a_2}_{A_2}\cdots\ket{a_N}_{A_N}$
(or $\ket{a_1a_2\cdots a_N}_{A_1A_2\cdots A_N}$). Similarly, we set
the sequential structure of Bob's space as $B_1B_2\cdots
B_NY_1Y_2\cdots Y_N$, in other words, its basis vector has the form
$\ket{b_1}_{B_1}\ket{b_2}_{B_2}\cdots\ket{b_N}_{B_N}
\ket{k_1}_{Y_1}\ket{k_2}_{Y_2}\cdots\ket{k_N}_{Y_N}$. It is clear
that for a $N$-qubit system, its space structure can be represented
by a bit-string with the length of $N$.

In order to write our formula compactly and clearly, and then prove
our protocols more conveniently, we need to use some general
swapping transformations, for example, $F_N(i,j)$, $P_N(j,k)$,
$\Lambda(2,N)$, $\Omega(2,N)$, $\Omega(3,N)$ and $\Upsilon(3,N)$,
that are studied in Appendix A and we will not repeatedly write
their definitions here.

In addition, we still need to define
\beq\Theta_N=\Omega(3,N)\Upsilon^{-1}(3,N)\eeq and then introduce
\beq \Theta_A(n)=\left(I_{2^n}\otimes\Lambda(2,N)\otimes
I_{2^N}\right)\left(\Theta_n\otimes I_{2^{3N-2n)}}\right),\eeq \beq
\Theta_B(n)=\left(I_{2^n}\otimes\Upsilon(3,N)\right)\left(\Theta_n\otimes
I_{2^{3N-2n)}}\right),\eeq \beq \Theta_C(n)=\left(\Theta_n\otimes
I_{2^{3N-2n)}}\right).\eeq Thus, we have \beqa & &
\Theta_A(n)\left[\left(\bigotimes_{m=1}^n\ket{a_mb_mc_m}_{A_mB_mC_m}\right)
\left(\bigotimes_{s=n+1}^N\ket{a_sb_s}_{A_sB_s}\right)
\otimes\left(\bigotimes_{t=1}^N\ket{k_t}_{Y_t}\right)\right]
\nonumber\\
& &\quad = \left(\bigotimes_{m=1}^n\ket{c_m}_{C_m}\right)
\left(\bigotimes_{s=n+1}^N\ket{a_s}_{A_s}\right)
\left(\bigotimes_{m=1}^N\ket{b_m}_{B_m}\right)
\otimes\left(\bigotimes_{m=1}^N\ket{k_m}_{Y_m}\right), \eeqa \beqa &
&
\Theta_B(n)\left[\left(\bigotimes_{m=1}^n\ket{a_mb_mc_m}_{A_mB_mC_m}\right)
\left(\bigotimes_{s=n+1}^N\ket{a_sb_s}_{A_sB_s}\right)
\otimes\left(\bigotimes_{s=1}^N\ket{k_s}_{Y_s}\right)\right]
\nonumber\\
& &\quad = \left(\bigotimes_{m=1}^n\ket{c_m}_{C_m}\right)
\left(\bigotimes_{s=1}^N\ket{a_sb_sk_s}_{A_sB_sY_s}\right), \eeqa
\beqa & &
\Theta_C(n)\left[\left(\bigotimes_{m=1}^n\ket{a_mb_mc_m}_{A_mB_mC_m}\right)
\left(\bigotimes_{s=n+1}^N\ket{a_sb_s}_{A_sB_s}\right)
\otimes\left(\bigotimes_{t=1}^N\ket{k_t}_{Y_t}\right)\right]
\nonumber\\
& &\quad = \left(\bigotimes_{m=1}^n\ket{c_m}_{C_m}\right)
\left(\bigotimes_{s=1}^N\ket{a_sb_s}_{A_sB_s}\right)
\otimes\left(\bigotimes_{t=1}^N\ket{k_t}_{Y_t}\right). \eeqa
Similarly, we can obtain the transformed relations acting on the
operations (or matrices).

By comparing with the cases of one qubit, we can extend the
controlled and combined RIO protocols to the cases of $N$ qubits in
terms of our restricted sets. However, we find that the variety of
protocols is more obvious, the expressions and proofs of protocols
get a little complicated. Our protocols are still made up of seven
steps for controlled and combined remote implementations of
$N$-qubit quantum operations belonging to our restricted sets. For
the cases with $n< N$ controllers, we only need $n$ GHZ states and
$N-n$ Bell pairs. While when two senders are fetched in, we need $N$
GHZ states.

Without loss of generality, when with $n$ controllers, we set the
former $n$ shared entangled states as GHZ states, the initial state
reads \beq \ket{\Psi_N^{\rm ini}}=\left(\bigotimes_{m=1}^n\ket{\rm
GHZ^+}_{A_mB_mC_m}\right)\otimes
\left(\bigotimes_{s=n+1}^N\ket{\Phi^+}_{A_sB_s}\right)
\otimes\ket{\xi}_{Y_1Y_2\cdots Y_N}, \eeq when with two senders, we
take the $N$ shared GHZ states, the initial state becomes \beq
\ket{\Psi_N^{\rm ini}}=\left(\bigotimes_{m=1}^N\ket{\rm
GHZ^+}_{A_mB_mC_m}\right) \otimes\ket{\xi}_{Y_1Y_2\cdots Y_N}, \eeq
where $\ket{\xi}_{y_1Y_2\cdots y_N}$ is an arbitrary (unknown) pure
state in the $N$-qubit systems: \beq \ket{\xi}_{Y_1\cdots
Y_N}=\sum_{k_1,\cdots,k_N=0}^1 y_{k_1\cdots k_N}\ket{k_1k_2\cdots
k_N}.\eeq

Therefore, we know that the space structures are initially \beq
\prod_{m=1}^nA_mB_mC_m\prod_{s=n+1}^N A_sB_s\prod_{t=1}^NY_t\eeq for
the case with $n$ controllers, and \beq
\prod_{m=1}^NA_mB_mC_m\prod_{t=1}^NY_t\eeq for the case with two
senders.

For simplicity, in the following, we only present the operations and
measurements done by the three partite subsystems and omit the steps
of classical communications. Of course, we still have to remember
the implementing sequence of them. Moreover, we do not detailedly
account for the significance and action of every step, which can be
understood from the cases of one qubit.

\subsection{With $n$ controllers}

For the cases with $n$ controllers, we set Charlies as controllers,
Alice as a sender and Bob as a receiver.

The controllers's (Charlies') startup is \beq
\mathcal{C}(c_1,\cdots,c_n)=\Theta_C^{-1}(n)\left[
\bigotimes_{m=1}^n\left(\ket{c_m}_{C_m}\bra{c_m}H^{C_m}\right)\otimes
I_{2^{3N}}\right]\Theta_C(n).\eeq

Bob's prior preparation is \beqa {\mathcal{P}}^{\rm
add}_B(c_1,c_2,\cdots,c_n)
&=&\Theta^{-1}_B(n)\left\{I_{2^n}\otimes\left(\bigotimes_{m=1}^n
\sigma_0^{A_m}\otimes\mathfrak{r}^{B_m}(c_m)\otimes\sigma_0^{Y_m}\right)\otimes
I_{2^{3(N-n)}}\right\}\Theta_B(n).\eeqa

Bob's preparing is \beqa \label{NqBp}
{\mathcal{P}}_B(b_1,b_2,\cdots,b_N)&=&\Theta^{-1}_B(n)
\left\{I_{2^{n}}\otimes\left[\bigotimes_{m=1}^N
\sigma_0^{A_m}\otimes
\left(\ket{b_m}_{B_m}\bra{b_m}\otimes\sigma_0^{Y_m}\right)\right.\right.\nonumber\\
& &\left.\left.
\times\left(\sigma_0^{B_m}\otimes\ket{0}_{Y_m}\bra{0}
+\sigma_1^{B_m}\otimes\ket{1}_{Y_m}\bra{1}\right)\right]\right\}\Theta_B(n).
\eeqa

Alice's prior sending is \beqa {\mathcal{S}}_A^{\rm
add}(c_1,c_2,\cdots,c_n)
&=&\Theta^{-1}_A(n)\left\{I_{2^n}\otimes\left(\bigotimes_{m=1}^n
\mathfrak{r}(c_m)\right)\otimes I_{2^{N-n}} \otimes
I_{4^N}\right\}\Theta_A(n).\eeqa

Alice's sending is \beqa \label{NqAs}
{\mathcal{S}}_A(a_1,b_1,a_2,b_2,\cdots,a_N,b_N;x,u)
&=&\Theta^{-1}_A(n)\left\{I_{2^n}\otimes\left[\left(\bigotimes_{m=1}^N
\ket{a_m}_{A_m}\bra{a_m}\right)\left(\bigotimes_{m=1}^N
H^{A_m}\right)\right.\right.\nonumber\\ & & \left.\times T^r_N(x,u)
\left.
\left(\bigotimes_{m=1}^N\sigma_{b_m}^{A_m}\right)\right]\otimes
I_{4^N}\right\}\Theta_A(n).\eeqa

Bob's supplementary recovering is \beq {\mathcal{R}}_B^{\rm
add}(c_1,c_2\cdots
c_n)=\left(\bigotimes_{m=1}^n\sigma_0^{A_m}\otimes
\mathfrak{r}(c_m)\otimes\sigma_0^{C_m}\right)\otimes
I_{2^{2(N-n)}}\otimes \left(\bigotimes_{m=1}^n
\mathfrak{r}(c_m)^{Y_m}\right)\otimes I_{2^{N-n}}. \eeq

Bob's recovering is \beq {\mathcal{R}}_B(a_1,a_2\cdots
a_N;x)=I_{2^{(2N+n)}}\otimes\left(\bigotimes_{m=1}^N
\mathfrak{r}(a_m)^{Y_m}\right){R}_N(x).\eeq

In particular, since ${R}_N(x)\left[\left(\bigotimes_{m=1}^n
\mathfrak{r}(c_m)^{Y_m}\right)\otimes
I_{2^{N-n}}\right]R^\dagger_N(x)$ and $\left(\bigotimes_{m=1}^N
\mathfrak{r}(a_m)^{Y_m}\right)$ are diagonal, they commute each
other. Again from $R^\dagger_N(x)R_N(x)=I_{2^{N}}$, it follows that
\beqa & & {R}_N(x)\left[\left(\bigotimes_{m=1}^n
\mathfrak{r}(c_m)^{Y_m}\right)\otimes
I_{2^{N-n}}\right]R^\dagger_N(x)\left[\left(\bigotimes_{m=1}^N
\mathfrak{r}(a_m)^{Y_m}\right){R}_N(x)\right]\nonumber\\
& &\quad =\left[\left(\bigotimes_{m=1}^N
\mathfrak{r}(a_m)^{Y_m}\right)
{R}_N(x)\right]\left[\left(\bigotimes_{m=1}^n
\mathfrak{r}(c_m)^{Y_m}\right)\otimes I_{2^{N-n}}\right]. \eeqa
Therefore, we obtain a final recovery operation \beqa
{\mathcal{R}}_B^{\rm aft}(c_1,c_2\cdots
c_n;x)&=&\left(\bigotimes_{m=1}^n\sigma_0^{A_m}\otimes
\mathfrak{r}(c_m)\otimes\sigma_0^{C_m}\right)\otimes
I_{2^{2(N-n)}}\nonumber\\
& &\otimes \left\{{R}_N(x)\left[\left(\bigotimes_{m=1}^n
\mathfrak{r}(c_m)^{Y_m}\right)\otimes
I_{2^{N-n}}\right]R^\dagger_N(x)\right\}.\eeqa Obviously, such a
finally additional recovery-operation is complicated in form
compared with the other additional operations. Perhaps, it is not
worth being used in our protocols.

It must be pointed out that three times of classical communication
are, respectively: (1) Charlies to Alice or Bob $n$ $c$-bits; (2)
Bob to Alice $N$ $c$-bits; (3) Alice to Bob
$N+\left[\log_2(2^N!)\right]+1$ $c$-bits. ($x$ may be encoded by
$\left[\log_2(2^N!)\right]+1$ $c$-bit string, where $[\cdots]$ means
taking the integer part).

Based on the kinds and time of the controllers distributing to the
sender or receiver, we only can use one beforehand operation for Bob
or Alice or Charlies, which has been seen in the cases of one qubit.
The whole operations and measurements can be jointly written
according to four cases (omitting arguments for simplicity):

(1) Alice (sender) obtains password before her sending \beq
\mathcal{I}_R(1)=\mathcal{R}_B\mathcal{S}_A\mathcal{S}_A^{\rm
add}\mathcal{P}_B;\eeq

(2) Bob (receiver) obtains password before his preparation \beq
\mathcal{I}_R(2)=\mathcal{R}_B\mathcal{S}_A\mathcal{P}_B\mathcal{P}_B^{\rm
add};\eeq

(3) Bob (receiver) obtains password after his preparation \beq
\mathcal{I}_R(3)=\mathcal{R}_B\mathcal{R}_B^{\rm
add}\mathcal{S}_A\mathcal{P}_B;\eeq

(4) Bob (receiver) obtains password after his recovery operations
\beq \mathcal{I}_R(4)=\mathcal{R}^{\rm
aft}\mathcal{R}_B\mathcal{S}_A\mathcal{P}_B.\eeq

\subsection{With two senders}

In the case of two senders, we set Bob as a receiver, Alice as the
first sender and Charlie as the second sender. It is different from
the cases of one qubit, where Charlie is a receiver, Alice is the
first sender and Bob is the second sender. However, it is
unimportant since the symmetry among three partite subsystems.

Bob's preparation and Alice's sending are the same as the above
(\ref{NqBp}) and (\ref{NqAs}), but $n=N$. Charlie's second sending
is \beqa \label{NqCs} {\mathcal{S}}_C(c_1,c_2,\cdots,c_N;x,y,v)&=&
\Theta^{-1}_C(N)\left\{\left[\left(\bigotimes_{m=1}^N\ket{c_m}_{C_m}\bra{c_m}
\right)\left(\bigotimes_{m=1}^N H^{C_m}\right)\right.\right.\nonumber\\
& & \left.\left. \times T_N^r(y,v)R_N(x)\right]\otimes
I_{2^{3N}}\right\}\Theta_C(N).\eeqa

Bob's recovery operation is \beqa & &
{\mathcal{R}}_B(a_1,a_2,\cdots,a_N,b_1,b_2,\cdots,b_n,c_1,c_2,\cdots,c_N;x,y)\nonumber\\
& &\quad =I_{2^{3N}} \otimes\left[\left(\bigotimes_{m=1}^N
\mathfrak{r}^{Y_m}(c_m)\right) R_N(y)\right]
\left[\bigotimes_{m=1}^N\left(\mathfrak{r}^{Y_m}(a_m)\right)
R_N(x)\right]\left(\bigotimes_{m=1}^N\sigma_{b_m}^{Y_m}\right),\eeqa
where $\mathfrak{r}(y)$ is defined by Eq. (\ref{rpg}).

It must be pointed out that four times of classical communication
are respectively: (1) Bob to Alice $N$ $c$-bits and to Charlies $N$
$c$-bits; (2) Alice to Bob $N+\left[\log_2(2^N!)\right]+1$ $c$-bits;
(3) Alice to Charlie $\left[\log_2(2^N!)\right]+1$ $c$-bits; (4)
Charlie to Bob $N+\left[\log_2(2^N!)\right]+1$ $c$-bits.

Obviously, the whole operations and measurements can be jointly
written as \beq
\mathcal{I}_R=\mathcal{R}_B\mathcal{S}_C\mathcal{S}_A\mathcal{P}_B.
\eeq

It is not difficult to prove our protocols for the cases of $N$
qubits because all of steps are similar to the cases of one qubit,
which is put in Appendix B.

\section{Discussion and conclusion}\label{sec6}

We have investigated the controlled and combined remote
implementation of the quantum operations belonging to our restricted
sets \cite{MyRIO} using GHZ state(s). The main motivations to use
GHZ state(s) in our protocols are to enhance security, increase
variety, extend applications as well as advance efficiency via
fetching in many controllers and two senders.

It must be emphasized that knowing the forms of the restricted sets
of quantum operations that can be remotely implemented is a key
matter to successfully carry out the RIO protocols. In our resent
work \cite{MyRIO}, we obtained their general and explicit forms.
Moreover, we provided evidence of the uniqueness and optimization of
our restricted sets based on the precondition that our protocol only
uses $N$ maximally entangled states. It must be emphasized that
before the beginning of our protocols, we have to build two mapping
tables, one of them provides one-to-one mapping from $T^r_N(x)\in
\mathbb{T}^r_N$ to the classical information $x$ which is known by
the senders, the other one provides one-to-one mapping from a
classical information $x$ to $R_N(x)$ which is known by the
receiver. Since the unified recovery operations are obtained, all of
quantum operations belonging to our restricted sets can be remotely
implemented via our protocols in a faithful and determined way. In
addition, although the important and interesting quantum operations
belonging to the restricted sets should be unitary, but this
limitation does not affect our protocol.

In this paper, we not only propose our protocols in detail, but also
prove them strictly in the cases of one and more than one qubit.
Through respectively describing the cases with the one or many
controllers as well as with one or two senders, we explain clearly
their roles in our protocols.

It should be pointed out that the implementations of $R_N(x)$ are
important in our protocols. It is a key to design a recovery quantum
circuits in the near future. In principle, we can construct $R_N(x)$
by using a series of universal gates \cite{NielsenBook}. In special,
we have found that $R_2(x)$ can be constructed by $C^{\rm not}$ and
$\sigma_{x}$ \cite{Ours2}:
\begin{eqnarray}
R_{2}(1)&=&I_{Y_{1}}\otimes I_{Y_{2}},\\
R_{2}(2)&=&C^{\rm not}(Y_{1},Y_{2}),\\
R_{2}(3)&=&C^{\rm not}(Y_{2},Y_{1}) C_{\sc not}(Y_{1},Y_{2})
C^{\rm not}(Y_{2},Y_{1})\\
R_{2}(4)&=&C^{\rm not}(Y_{2},Y_{1})
C^{\rm not}(Y_{1},Y_{2}),\\
R_{2}(5)&=&C^{\rm not}(Y_{1},Y_{2}) C^{\rm not}(Y_{2},Y_{1}),\\
R_{2}(6)&=&C^{\rm not}(Y_{2},Y_{1}),\\
R_{2}(7)&=&C^{\rm not}(Y_{1},Y_{2})(I\otimes\sigma_{1}),\\
R_{2}(8)&=&(I\otimes\sigma_{1}),\\
R_{2}(9)&=&(\sigma_{1}\otimes I) C^{\rm not}(Y_{1},Y_{2})
C^{\rm not}(Y_{2},Y_{1}),\\
R_{2}(10)&=&C^{\rm not}(Y_{2},Y_{1})(I\otimes\sigma_{1}),\\
R_{2}(11)&=&C^{\rm not}(Y_{2},Y_{1})(\sigma_{1}\otimes I)
C^{\rm not}(Y_{1},Y_{2}) C^{\rm not}(Y_{2},Y_{1}),\\
R_{2}(12)&=&C^{\rm not}(Y_{2},Y_{1})
C^{\rm not}(Y_{1},Y_{2})(I\otimes\sigma_{1}),\\
R_{2}(13)&=&C^{\rm not}(Y_{2},Y_{1})
C^{\rm not}(Y_{1},Y_{2})(\sigma_{1}\otimes I),\\
R_{2}(14)&=&C^{\rm not}(Y_{2},Y_{1}) C^{\rm
not}(Y_{1},Y_{2})(\sigma_{1}\otimes I)
 C^{\rm not}(Y_{2},Y_{1}),\\
R_{2}(15)&=&C^{\rm not}(Y_{2},Y_{1})(\sigma_{1}\otimes I),\\
R_{2}(16)&=&C^{\rm not}(Y_{1},Y_{2})(\sigma_{1}\otimes I) C^{\rm
not}(Y_{2},Y_{1}),\\
R_{2}(17)&=&(\sigma_{1}\otimes I),\\
R_{2}(18)&=&C^{\rm not}(Y_{1},Y_{2}) (\sigma_{1}\otimes I),\\
R_{2}(19)&=&(I\otimes\sigma_{1})
C^{\rm not}(Y_{2},Y_{1}),\\
R_{2}(20)&=&C^{\rm not}(Y_{1},Y_{2})(I\otimes \sigma_{1})
 C^{\rm not}(Y_{2},Y_{1}),\\
R_{2}(21)&=&C^{\rm not}(Y_{2},Y_{1}) (\sigma_{1}\otimes I)
C^{\rm not}(Y_{1},Y_{2}),\\
R_{2}(22)&=&C^{\rm not}(Y_{2},Y_{1}) C^{\rm not}(Y_{1},Y_{2})
(I\otimes\sigma_{1}) C^{\rm not}(Y_{2},Y_{1}),\\
R_{2}(23)&=&(\sigma_{1}\otimes I) C^{\rm not}(Y_{1},Y_{2}),\\
R_{2}(24)&=&(\sigma_{1}\otimes\sigma_{1}).
\end{eqnarray}
where $C^{\rm not}(Y_{1},Y_{2})$ means that we use qubit $Y_{1}$ as
the control qubit, $Y_{2}$ as the target qubit to do the
control-{\sc not} transformation, and $C^{\rm not}(Y_{2},Y_{1})$
means we use qubit $Y_{2}$ as the control qubit and qubit $Y_{1}$ as
the target qubit. Furthermore, we are interesting in the
construction of a unified recovery quantum circuit, which will be
studied in our other manuscript. It is worthy pointing out that the
unified recovery operations in our protocols imply that quantum
operations that can be remotely implemented can belong to all of the
restricted sets but not only a kind of restricted set. This
advantage obviously reveals that the power of remote implementations
of quantum operations in our protocols is enhanced.

Form the controlled RIOs, we have seen that the controller(s) is
(are) an (a group of) administrator(s) in our protocols. If the
controllers (controller) accept(s) the application of remote
implementations of operations from sender and receiver, or intend(s)
to let sender(s) and receiver(s) carry out the RIO task, they
(he/she) will perform the startup operation (controlling step) and
then transfer the classical information as a ``password" to the
sender(s) (allowing step) or the receiver so that the protocols can
begin and be faithfully completed. When controllers, a sender and a
receiver share $N$ GHZ states, it does not mean that the sender and
receiver can carry out RIO protocols. This is because that the
quantum channel between the sender and receiver has not been opened.
The startup of the quantum channel is obtained by the controllers'
operation. It is just one of reasons why we use the name of
controller(s). Then, the controller(s) transfers his/her classical
information as a ``password". However, as soon as the password is
transferred, the controller has no any means to stop the protocols.
Therefore, we suggest a scheme to delay this transmission (password
distributing) and send the password(s) to the receiver until the
finishing of the receiver's standard recovered step so that the
controller(s) keeps his/her's interrupting right up to the end of
the protocols, that is, ``saying last word".  However, it is
possible to lead to a little complicated form of the receiver's
recovery operation for the multiqubits cases, so we may give up this
kind of scheme and put the additional recovery operation before the
standard recovery operations.

It should be pointed out that when three partite subsystems share
$N$ GHZ states, their position and right are symmetric. Therefore,
any partite subsystem can be one of controller, sender and receiver.
The controller is determined by the other two parties' choice based
on their requirement of RIO, and/or his/her own decision in order to
authorize the other two partite subsystems carrying out RIO. If an
advanced administrator nominates a controller, he/she can demand
this controller to open the quantum channel between the other two
partite subsystems but keep the classical information in hands as a
controlled means. If the number of shared GHZ states is $n$ less
than $N$, only two partite subsystems will be symmetric and they can
choose as either of a sender and a receiver, the other one partite
subsystem with $n$ qubits only can play the controllers.

For the combined remote implementation of quantum operations, we
also have displayed that two senders respectively complete the
remote implementations of two parts of a quantum operation and then
combine them together to obtain the finally remote implementation of
the whole quantum operations via GHZ states. Because the second
sender has to know the classical information from the first sender,
the combination of two parts of operations has a sequence.  This
implies that the cooperation of all the senders are needed. It is
clear that the security of remote implementations of operations is
enhanced. The related reasons have been stated in the introduction.
This advantage does not exist in the RIO protocol with two senders
using two Bell states. In practice, it is possible that different
senders have different operational resources and different
operational rights, therefore, we can set a suitable combination of
their resource and right. It implies that the combined remote
implementation can overcome the senders's possible shortcoming and
help us to farthest arrive at the power of our protocols in theory.
In addition, it is interesting to study the quantum resource cost in
the RIO protocol with two senders by comparing the different schemes
using one GHZ state with using two Bell state.

Furthermore, if we wish to consider our protocols with more than two
senders, or both many controllers and many senders, then the
entangled states of three partite subsystems are not enough when
only using $N$ GHZ states in order to remotely implement the
operations of $N$ qubits. In general, for the cases of quantum
operations of $N$ qubits,  if there are $n\;(\leq N))$ controllers
and $m\; (\leq N))$ senders, we need using $N$ EPR-GHZ states at
least with $m+n+1$ partite subsystems.

In summary, using $GHZ$ states in the RIOs protocols indeed can
enhance their security, increase their variety, extend their
possible applications, and even advance their efficiency. These
advantages can not be replaced by using Bell states. Therefore, we
can say the different quantum resources have the different features
and purposes in quantum information processing and communications.

\section*{Acknowledgments}

We are grateful all the collaborators of our quantum theory group in
the institute for theoretical physics of our university. This work
was funded by the National Fundamental Research Program of China
under No. 2001CB309310, partially supported by the National Natural
Science Foundation of China under Grant No. 60573008.

\begin{appendix}

\section{Swapping transformation}

In this appendix, we first study the general swapping
transformations, which are the combinations of a series of usual
swapping transformations. They are used in our protocols in order to
express our formula clearly and compactly, and prove our protocols
more easily.

Note that a swapping transformation of two neighbor qubits is
defined by \beq S_W=\left(\begin{array}{cccc}
1 & 0 & 0 &0\\
0& 0 & 1 & 0\\
0& 1 & 0 & 0\\
0& 0 & 0 & 1
\end{array}\right).
\eeq Its action is \beq
S_W\ket{\alpha_X\beta_Y}=\ket{\beta_Y\alpha_X},\quad S_W(M^X\otimes
M^Y)S_W=M^Y\otimes M^X. \eeq This means that the swapping
transformation changes the space structure $H_X\otimes H_Y$ into
$H_Y\otimes H_X$.

For an $N$-qubit system, the swapping gate of the $i$th qubit and
the $(i+1)$th qubit reads \beq S_N(i,i+1)=
\sigma_0^{\otimes(i-1)}\otimes S_W\otimes\sigma_0^{\otimes(N-i-1)}.
\eeq Two rearranged transformations are defined by
\begin{equation}
F_N(i, j)=\prod_{\alpha=1\leftarrow}^{j-i}S_N(j-\alpha,j+1-\alpha)
\end{equation}
\begin{equation}
P_N(j, k)=\prod_{\beta=j\leftarrow}^{k-1}S_N(\beta,\beta+1)
\end{equation}
where $F_N(i,j)$ extracts out the spin-state of site $j$, and
rearranges it forwards to the site $i$ ($i<j$) in the qubit-string,
where $P_N(j,k)$ extracts out the spin-state of site $j$, and
rearranges it backwards to the site $k$ ($k>j$) in the qubit-string.
Note that ``$\leftarrow$" means that the factors are arranged from
right to left corresponding to $\alpha, \beta$ from small to large.
Now, in terms of $P(j,k)$, we can introduce two general swapping
transformations with the forms
\begin{equation}
{\Lambda}(2,N)=\prod_{i=1\leftarrow}^{N-1}P_{2N}\left(2(N-i),2N-i\right),\quad
(N\geq 2),
\end{equation}
\begin{equation}
{\Omega}(2,N)=\prod_{i=1\leftarrow}^{N}P_{2N}\left(1,2N\right),\quad
(N\geq 2).
\end{equation}
Thus, \beq {\Lambda}(2,N)\left(\bigotimes_{i=1}^N\ket{a_i
b_i}\right)=\left(\bigotimes_{i=1}^N
\ket{a_i}\right)\otimes\left(\bigotimes_{j=1}^N
\ket{b_j}\right),\eeq \beq {\Lambda}(2,N)\left(\bigotimes_{k=1}^N
\left(M_{\alpha_i}^{A_i}\otimes
M_{\beta_i}^{B_i}\right)\right){\Lambda}^{-1}(2,N)=\left(\bigotimes_{i=1}^N
M_{\alpha_i}^{A_i}\right)\otimes\left(\bigotimes_{j=1}^N
M_{\beta_j}^{B_j}\right), \eeq \beq
{\Omega}(2,N)\left[\left(\bigotimes_{i=1}^N\ket{a_i
}\right)\otimes\left(\bigotimes_{j=1}^N\ket{b_j
}\right)\right]=\left(\bigotimes_{i=1}^N
\ket{b_i}\right)\otimes\left(\bigotimes_{j=1}^N
\ket{a_j}\right),\eeq \beq
{\Omega}(2,N)\left[\left(\bigotimes_{i=1}^N
M_{\alpha_i}^{A_i}\right)\left(\bigotimes_{i=1}^N
M_{\beta_i}^{B_i}\right)\right]{\Omega}^{-1}(2,N)=\left(\bigotimes_{i=1}^N
M_{\beta_i}^{B_i}\right)\otimes\left(\bigotimes_{j=1}^N
M_{\alpha_j}^{A_j}\right). \eeq Similarly, we can introduce \beqa
{\Upsilon}(3,N)=\prod_{i=1\leftarrow}^{N-1}F_{3N}\left(3i,2N+i\right),\quad
(N\geq 2); \\
{\Upsilon}(4,N)=\prod_{i=1\leftarrow}^{N-1}F_{4N}\left(4i,3N+i\right),\quad
(N\geq 2). \eeqa \beq \Gamma(3,N)=\left(I_{2^N}\otimes
\Omega(2,N)\right)\left(\Lambda(2,N)\otimes I_{2^N}\right).\eeq
Thus, \beq {\Upsilon}(3,N)\left(\bigotimes_{i=1}^N\ket{a_i
b_i}\right)\otimes\left(\bigotimes_{j=1}^N\ket{k_j}\right)=\bigotimes_{i=1}^N
\ket{a_i b_i k_i},\eeq \beq {\Upsilon}(3,N)\left[\bigotimes_{k=1}^N
\left(M_{\alpha_i}^{A_i}\otimes
M_{\beta_i}^{B_i}\right)\right]\left(\bigotimes_{j=1}^N
M_{\gamma_j}^{Y_j}\right){\Upsilon}^{-1}(3,N) =\bigotimes_{i=1}^N
\left(M_{\alpha_i}^{A_i}\otimes M_{\beta_i}^{B_i}\otimes
M_{\gamma_i}^{Y_i}\right). \eeq
\beq {\Upsilon}(4,N)\left(\bigotimes_{i=1}^N\ket{a_i
b_ic_i}\right)\otimes\left(\bigotimes_{j=1}^N\ket{k_j}\right)=\bigotimes_{i=1}^N
\ket{a_i b_i c_ik_i},\eeq \beq
{\Upsilon}(4,N)\left[\bigotimes_{k=1}^N
\left(M_{\alpha_i}^{A_i}\otimes M_{\beta_i}^{B_i}\otimes
M_{\gamma_i}^{C_i}\right)\right]\left(\bigotimes_{j=1}^N
M_{\delta_j}^{Y_j}\right){\Upsilon}^{-1}(4,N) =\bigotimes_{i=1}^N
\left(M_{\alpha_i}^{A_i}\otimes M_{\beta_i}^{B_i}\otimes
M_{\gamma_i}^{C_i}\otimes M_{\delta_i}^{Y_i}\right). \eeq
\beq {\Gamma}(3,N)\left(\bigotimes_{i=1}^N\ket{a_i
b_i}\right)\otimes\left(\bigotimes_{j=1}^N\ket{k_j}\right)=\left(\bigotimes_{i=1}^N
\ket{a_i}\right)\otimes\left(\bigotimes_{j=1}^N\ket{
k_j}\right)\otimes\left(\bigotimes_{k=1}^N \ket{b_k}\right),\eeq
\beqa & &{\Gamma}(3,N)\left[\bigotimes_{k=1}^N
\left(M_{\alpha_i}^{A_i}\otimes
M_{\beta_i}^{B_i}\right)\right]\left(\bigotimes_{j=1}^N
M_{\gamma_j}^{Y_j}\right){\Gamma}^{-1}(3,N)\nonumber\\
& &\quad =\left(\bigotimes_{i=1}^N
M_{\alpha_i}^{A_i}\right)\otimes\left(\bigotimes_{j=1}^N
M_{\gamma_j}^{Y_j}\right)\otimes\left(\bigotimes_{k=1}^N
M_{\beta_k}^{B_k}\right). \eeqa

For the cases with $n$ GHZ states and $N-n$ Bell states, we
introduce \beq \Omega(3,N)=\left(\Omega(2,N)\otimes
I_{2^N}\right)\left(I_{2^N}\otimes\Omega(2,N)\right).\eeq Obviously
\beq {\Omega}(3,N)\left[\left(\bigotimes_{i=1}^N\ket{a_i b_i
}\right)\otimes\left(\bigotimes_{j=1}^N\ket{c_j
}\right)\right]=\left(\bigotimes_{i=1}^N
\ket{c_i}\right)\otimes\left(\bigotimes_{j=1}^N
\ket{a_jb_j}\right),\eeq \beq
{\Omega}(3,N)\left\{\left[\bigotimes_{i=1}^N\left(
M_{\alpha_i}^{A_i}\otimes
M_{\beta_i}^{B_i}\right)\right]\left(\bigotimes_{i=1}^N
M_{\gamma_i}^{C_i}\right)\right\}\Omega^{-1}(3,N)=\left(\bigotimes_{i=1}^N
M_{\gamma_i}^{C_i}\right)\otimes\left[\bigotimes_{j=1}^N\left(
M_{\alpha_j}^{A_j}\otimes M_{\beta_j}^{B_j}\right)\right]. \eeq

More generally, consider the set $\mathbb{Q}_N$ to be a whole
permutation of the bit-string ${a_1a_2\cdots a_N}$, and denote the
$z$ element with a bit-string form $Q(z)=q_1(z)q_2(z)\cdots q_N(z)$,
we always can obtain such a general swapping transformation $W_N$
that a computational basis $\ket{a_1a_2\cdots a_N}$ of $N$-qubit
systems can be swapped as another basis $\ket{q_1(z)q_2(z)\cdots
q_N(z)}$ in which $q_1(z)q_2(z)\cdots q_N(z)$ is an arbitrary
element of $\mathbb{Q}_N$. Thus, we can write a given general
swapping transformation $W_N\left(a_1a_2\cdots a_N\rightarrow
q_1(z)q_2(z)\cdots q_N(z)\right)$,
\beq W_N\left[{a_1a_2\cdots
a_N}\rightarrow {q_1(z)q_2(z)\cdots q_N(z)}\right]\ket{a_1a_2\cdots
a_N}=\ket{q_1(z)q_2(z)\cdots q_N(z)}.\eeq
Furthermore, if we denote
two dimensional space $A_i$ spanned by $\ket{a_i}$ ($a_i=0,1$ and
$i=1,2,\cdots N$), while $M^{A_i}$ is a matrix belonging to this
space, we obviously have \beqa & &W_N^{-1}\left[{a_1a_2\cdots
a_N}\rightarrow {q_1(z)q_2(z)\cdots
q_N(z)}\right]\left(\prod_{i=1}^NM^{A_i}\right)
W_N\left[{a_1a_2\cdots a_N}\rightarrow {q_1(z)q_2(z)\cdots
q_N(z)}\right]=\left(\prod_{i=1}^NM^{A_{q_i(z)}}\right).\hskip
1.0cm\eeqa Therefore, the general swapping transformation $W_N$
defined above can be used to change the space structure of
multiqubits systems.

\section{The proof of our protocol in the cases more than one
qubit}

Here, we would like to prove our protocols of controlled and
combined RIO belonging to our restricted sets in the cases of more
than one qubit.

For the cases with $n$ controllers, since \beq
\left(I_4\otimes\ket{c}\bra{c}H\right)\ket{\rm
GHZ_+}=\frac{1}{2}\left(\ket{00}+(-1)^c\ket{11}\right)\otimes\ket{c}.\eeq
The initial state is transformed as \beqa
\ket{\Psi^{C}_N}&=&\mathcal{C}(c_1,c_2,\cdots,c_n)\ket{\Psi^{\rm
ini}_N}= \frac{1}{\sqrt{2^n}}\left(\bigotimes_{m=1}^n\ket{\rm
Bell_{c_m}}_{A_mB_m}
\otimes\ket{c_m}\right)\left(\bigotimes_{s=n+1}^N\ket{\rm
Bell_+}_{A_jB_j}\right)\otimes\ket{\xi}_{y_1\cdots y_N},\eeqa where
\beq \ket{\rm
Bell_{c_i}}=\frac{1}{\sqrt{2}}\left(\ket{00}+(-1)^{c_i}\ket{11}\right).\eeq
Again introducing the swapping transformation \beq
\Xi_N(n)=\left[I_n\otimes\Upsilon(3,N)\right]\Theta^{-1}_C(n),\eeq
where $\Upsilon(3,N)$ and $\Theta_C(n)$ is defined as above, we can
rewrite \beqa \ket{\Psi^{C}_N}&=&\frac{1}{\sqrt{2^{N+n}}}
 \Xi_N(n)\left(\bigotimes_{i=1}^n\ket{c_i}\right)\sum_{k_1,\cdots
k_N=0}^1 y_{k_1\cdots
k_N}\nonumber\\
& &\bigotimes_{m=1}^n\left(\mathfrak{r}(c_i)\otimes
I_4\right)\left(\ket{00k_m}_{A_mB_my_m}+(-1)^{c_m}\ket{11k_m}_{A_mB_my_m}\right)\\
&
&\bigotimes_{s=n+1}^N\left(\ket{00k_s}_{A_sB_sY_s}+\ket{11k_s}_{A_sB_sY_s}\right)\nonumber\\
&=&\frac{1}{\sqrt{2^{N+n}}}
 \Xi_N(n)\left(\bigotimes_{i=1}^n\ket{c_i}\right)\sum_{k_1,\cdots
k_N=0}^1 y_{k_1\cdots
k_N}\nonumber\\
& &\bigotimes_{m=1}^n\left(I_2\otimes\mathfrak{r}(c_i)\otimes
I_2\right)\left(\ket{00k_m}_{A_mB_my_m}+(-1)^{c_m}\ket{11k_m}_{A_mB_my_m}\right)\\
&
&\bigotimes_{s=n+1}^N\left(\ket{00k_s}_{A_sB_sY_s}+\ket{11k_s}_{A_sB_sY_s}\right).
\eeqa Since $\mathfrak{r}(c_i)\mathfrak{r}(c_i)=I_2$, therefore,
whatever Charlies transfer their information to Alice or Bob, all
factors $\mathfrak{r}(c_i)$ will be eliminated because the product
of it and the prior transformation $\mathfrak{r}(c_i)$ gets $1$. If
we delay the controllers' information to after the preparing, then
we can similarly discuss in terms of Eq. (\ref{rccnot}). If we delay
the controllers' information to the end of our protocols (that is
``say last word") in the cases of multiqubits, we will pay the price
that a more complicated additionally recovery operation is resulted
in. In the following, for simplicity, we only prove the case that
Charlies' information is transferred to Bob. The other kinds of our
protocols can be proved similar to the cases of one qubit.

From Bob's preparing (after his prior operation), it follows that
\beqa \ket{\Psi^P(b_1,\cdots
b_N)}&=&\mathcal{P}_B(b_1,b_2,\cdots,b_N)
\mathcal{P}_B^{\rm add}(c_1,c_2,\cdots,c_n)\ket{\Psi^{C}_N}\nonumber\\
&=&\frac{1}{\sqrt{2^{N+n}}}\Xi_N(n)\left(\bigotimes_{m=1}^n\ket{c_m}\right)\sum_{k_1,\cdots
k_N=0}^1 y_{k_1\cdots k_N}\bigotimes_{s=1}^N \nonumber\\
& & \times
\left[\sigma_0\left(\otimes\ket{b_s}\bra{b_s}\otimes\sigma_0\right)C^{\rm
not}(2,1)\right]\left(\ket{00k_s}_{A_sB_sY_s}+\ket{11k_s}_{A_sB_sY_s}\right)\nonumber\\
&=&\frac{1}{\sqrt{2^{N+n}}}\Xi_N(n)\left(\bigotimes_{m=1}^n\ket{c_m}\right)
\left[\bigotimes_{s=1}^N
\left(\sigma_{b_s}\otimes\sigma_0\right)\otimes
I_{2^N}\right]\nonumber\\
& &\times\sum_{k_1,\cdots k_N=0}^1 y_{k_1\cdots
k_N}\bigotimes_{t=1}^N\ket{k_tb_tk_t}_{A_tB_tY_t}, \eeqa where we
have used Eq. (\ref{pbell})

Actually, the physical idea to design our protocol is to perfectly
prepare the state of joint system being in the correlated
superposition. If there is the controller(s), the whole preparing is
completed by the controller's startup, receiver's setting and
sender's assistance, that
$\mathcal{P}=\mathcal{P}_S\mathcal{P}_B\mathcal{P}^{\rm
add}_B\mathcal{C}$. It is easy to see that \beqa
\label{psjs}\ket{\Psi_f^{\rm P}}&=&\mathcal{P}_S\ket{\Psi^{\rm
ini}}=\frac{1}{\sqrt{2^{N+n}}}\Xi_N(n)\left(\bigotimes_{m=1}^n\ket{c_m}\right)\otimes
\sum_{k_1,\cdots k_N=0}^1 y_{k_1\cdots
k_N}\bigotimes_{s=1}^N\ket{k_sb_sk_s}_{AsB_sy_s}\nonumber\\
&=&\frac{1}{\sqrt{2^{N+n}}}\Xi^\prime_N(n)\left(\bigotimes_{m=1}^n\ket{c_m}\right)\otimes
\sum_{k_1,\cdots k_N=0}^1 y_{k_1\cdots
k_N}\left(\bigotimes_{s=1}^N\ket{k_s}_{A_s}\right)\nonumber\\
& &\otimes\left(\bigotimes_{s=1}^N\ket{k_s}_{Y_s}\right)
\left(\bigotimes_{t=1}^N\ket{b_t}_{B_t}\right), \eeqa where we have
defined \beq
{\Xi}^\prime_N(n)=\Xi_N(n)\left[I_{2^n}\otimes\Upsilon^{-1}(3,N)\Gamma^{-1}(3,N)\right]
.\eeq

For simplicity, we only need to consider the subspace $A_1A_2\cdots
A_NY_1Y_2\cdots Y_N$, and omit the general swapping transformation
as well as the coefficient, we rewrite \beq \ket{\psi_f^{\rm
P}}\propto \sum_{k_1,\cdots k_N=0}^1 y_{k_1\cdots k_N}
\left(\bigotimes_{m=1}^N\ket{k_m}_{A_m}\right)
\otimes\left(\bigotimes_{m=1}^N\ket{k_m}_{Y_m}\right).\eeq

Thus, Alice's sending step and Bob's recovery operations yield the
final state in our interesting subsystem as \beqa \label
{nqfs1}\ket{\psi^{\rm final}_N}&\propto&
\bigotimes_{m=1}^N\ket{a_m}_{A_m}\otimes\sum_{k_1,\cdots
k_N=0}^1 y_{k_1\cdots k_N}\nonumber\\
& & \times
\left[\left(\bigotimes_{m=1}^N\bra{a_m}\right)\left(\bigotimes_{m=1}^N
H^{A_m}\right)T_N^r(x,t)\left(\bigotimes_{m=1}^N
\ket{k_m}\right)\right]\nonumber\\
& & \times
\left(\bigotimes_{m=1}^N\mathfrak{r}^{Y_m}(a_m)\right)R_N(x)
\left(\bigotimes_{m=1}^N\ket{k_m}_{Y_m}\right).\eeqa It is a key
matter that we can prove the relation \beq \label{ppr1}
T_N^r(1)R_N(x)=\sum_{m=1}^{2^N}
t_m\ket{i,D}\bra{m,D}\sum_{n=1}^{2^N}\ket{n,D}\bra{p_n(x),D}
=\sum_{m=1}^{2^N}t_m\ket{m,D}\bra{p_m(x),D}=T_N^r(x),\eeq and we
have known that \beq \label{tn1}
T^r_N(1)=\sum_{j_1,\cdots,j_N=0}^{1}t_{j_1j_2\cdots
j_N}\ket{j_1j_2\cdots j_N}\bra{j_1j_2\cdots j_N},\eeq \beq
\label{rn} \mathfrak{r}(a_m)=\sum_{l_m=0}^1(-1)^{a_m
l_m}\ket{l_m}\bra{l_m}.\eeq Substituting them into (\ref{nqfs1}), we
have \beqa \label{nqfs2}\ket{\psi^{\rm final}_N}&\propto&
\bigotimes_{m=1}^N\ket{a_m}_{A_m}\otimes\left\{\sum_{j_1,\cdots,j_N}^1
\sum_{k_1,\cdots k_N=0}^1\sum_{l_1,\cdots,l_N}^1 t_{j_1\cdots j_N}
y_{k_1\cdots
k_N}\right.\nonumber\\
& &\times\left(\prod_{m=1}^N\bra{a_m}H\ket{j_m}\right)
\left[\left(\bigotimes_{m=1}^N\bra{j_m}\right)R_N(x)
\left(\bigotimes_{m=1}^N\ket{k_m}_{Y_m}\right)\right]\nonumber\\
& &\times\left[\left(\bigotimes_{m=1}^N\bra{l_m}\right)R_N(x)
\left(\bigotimes_{m=1}^N\ket{k_m}_{Y_m}\right)\right]
\left(\prod_{m=1}^N(-1)^{a_m
l_m}\right)\bigotimes_{m=1}^N\ket{l_m}_{Y_m}.\eeqa Because that
$R_N(x)$ is such a matrix that its every row and every column only
has a nonzero element, we can obtain \beqa \label{rnor}&
&\left[\left(\bigotimes_{m=1}^N\bra{j_m}\right)R_N(x)
\left(\bigotimes_{m=1}^N\ket{k_m}_{Y_m}\right)\right]\left[
\left(\bigotimes_{m=1}^N\bra{l_m}\right)R_N(x)
\left(\bigotimes_{m=1}^N\ket{k_m}_{Y_m}\right)\right]
\nonumber\\
& & =\left(\prod_{m=1}^N\delta_{j_m l_m}\right)
\left[\left(\bigotimes_{m=1}^N\bra{j_m}\right)R_N(x)
\left(\bigotimes_{m=1}^N\ket{k_m}_{Y_m}\right)\right]. \eeqa Again
from \beq\label{hrr} \bra{a_m}H\ket{j_m}(-1)^{a_m
j_m}=\frac{1}{\sqrt{2}},\eeq we can derive out \beqa
\label{nqfs3}\ket{\psi^{\rm final}_N}&\propto&
\bigotimes_{m=1}^N\ket{a_m}_{A_m}\otimes\left\{\sum_{j_1,\cdots,j_N}^1\sum_{k_1,\cdots
k_N=0}^1\sum_{l_1,\cdots,l_N}^1 t_{j_1\cdots j_N} y_{k_1\cdots
k_N}\right.\nonumber\\
& &\times\left[\left(\bigotimes_{m=1}^N\bra{j_m}\right)R_N(x)
\left(\bigotimes_{m=1}^N\ket{k_m}_{Y_m}\right)\right]
\bigotimes_{m=1}^N\ket{j_m}_{Y_m}.\eeqa

If we directly act $T_N^r(x,t)$ on the unknown state, we have \beqa
T^r_N(x,t)\ket{\xi}_{Y_1\cdots Y_N}&=&\sum_{k_1,\cdots,k_N=0}^1
y_{k_1\cdots k_N}T^r_N(1,t)R(x)\ket{k_1k_2\cdots k_n}\nonumber\\
&=&\sum_{k_1,\cdots,k_N=0}^1\sum_{k_1,\cdots,k_N=0}^1 t_{j_1\cdots
j_N}y_{k_1\cdots k_N}\nonumber\\
& &\times\bra{j_1j_2\cdots j_N}R(x)\ket{k_1k_2\cdots
k_n}\ket{j_1j_2\cdots j_N}.\eeqa This means that

\beq \label{nqfs4}\ket{\psi^{\rm final}_N}\propto
\bigotimes_{i=1}^N\ket{a_i}_{A_i}\otimes \left(
T^r_N(x,t)\ket{\xi}_{y_1\cdots y_N}\right).\eeq

Finally, we add the other subspaces and restore the structure of
Hilbert's space by using the swapping transformations, and then
finish the proof of our protocols for the cases with $n$
controllers.

When there are two senders, our protocol is actually the combination
of twice remote implementations. However, after the first operation
is transferred remotely, the second sender's local system loses
perfectly correlation with the remote system to be operated. Since
we use one GHZ state as a quantum channel, the first transfer has
not exhausted all of correlation in the joint system. Compared with
the cases of one qubit, we can obtain the method to rebuild their
correlation through replacing $\sigma_d$ by the fixed form of the
first operation.

To our purpose, let us start with the state prepared by Bob: \beqa
\ket{\Psi^P_N}&=&\frac{1}{\sqrt{2^N}}\Upsilon^{-1}(4,N)\mathcal{P}_B(b_1,\cdots,
b_N)\Upsilon^{-1}(4,N)\left[\sum_{k_1,\cdots,k_N=0}^1y_{k_1\cdots
k_N}\bigotimes_{m=1}^N\left(\ket{000k_m}+\ket{111k_m}\right)\right]\nonumber\\
&=&
\frac{1}{\sqrt{2^N}}\Upsilon^{-1}(4,N)\left(\bigotimes_{m=1}F_4^{-1}(1,3)\right)\left\{\bigotimes_{m=1}^N
\left[I_4\otimes\left(\ket{b_m}\bra{b_m}\otimes\sigma_0\right)C^{\rm
not}(2,1)\right]\right.
\nonumber\\
&
&\left.\times\left[\ket{000k_m}_{C_mA_mB_mY_m}+\ket{111k_m}_{C_mA_mB_mY_m}\right]\right\}.
\eeqa Using Eq. (\ref{pghz}), we have \beqa \ket{\Psi^P_N}&=&
\frac{1}{\sqrt{2^N}}\Upsilon^{-1}(4,N)\left(\bigotimes_{m=1}F_4^{-1}(1,3)\right)\nonumber\\
& &\times\sum_{k_1\cdots k_N=0}^1y_{k_1\cdots
k_n}\left[\bigotimes_{m=1}^N\left(\sigma_{b_m}\otimes\sigma_{b_m}\otimes
I_4\right)\ket{k_mk_mb_mk_m}_{C_mA_mB_mY_m}\right]\nonumber\\
&=& \frac{1}{\sqrt{2^N}}\sum_{k_1\cdots k_N=0}^1y_{k_1\cdots
k_n}\left[\bigotimes_{m=1}^N\left(\sigma_{b_m}\otimes\sigma_0
\otimes\sigma_{b_m}\right)\ket{k_mb_mk_m}_{A_mB_mC_m}\right]
\otimes\left(\bigotimes_{m=1}^N\ket{k_m}_{Y_m}\right)\nonumber\\
&=& \frac{1}{\sqrt{2^N}}\left(\Lambda(2,N)\otimes
I_{4^N}\right)\left(\Gamma^{-1}(3,N)\otimes
I_{2^N}\right)\sum_{k_1\cdots k_N=0}^1y_{k_1\cdots
k_n}\left(\bigotimes_{m=1}^N\sigma_{b_m}\ket{k_m}_{A_m}\right)\nonumber\\
& &\otimes\left(\bigotimes_{m=1}^N\ket{b_m}_{B_m}\right)
\otimes\left(\bigotimes_{m=1}^N\sigma_{b_m}\ket{k_m}_{C_m}\right)
\otimes\left(\bigotimes_{m=1}^N\ket{k_m}_{Y_m}\right).\eeqa

Omitting the swapping transformations as well as coefficient and
keeping the relevant subspaces, we have \beq \ket{\psi^P_N}\propto
\sum_{k_1\cdots k_N=0}^1y_{k_1\cdots
k_n}\left(\bigotimes_{m=1}^N\sigma_{b_m}\ket{k_m}_{A_m}\right)
\otimes\left(\bigotimes_{m=1}^N\sigma_{b_m}\ket{k_m}_{C_m}\right)
\otimes\left(\bigotimes_{m=1}^N\ket{k_m}_{Y_m}\right).\eeq

Alice's sending, Charlie's sending and Bob's recovery operations
lead to \beqa \ket{\psi^{\rm final}_N}&\propto&
\left(\bigotimes_{m=1}^N\ket{a_m}_{A_m}\right)
\otimes\left(\bigotimes_{m=1}^N\ket{c_m}_{C_m}\right)
\nonumber\\
& &\times \sum_{k_1\cdots k_N=0}^1y_{k_1\cdots
k_n}\left[\left(\bigotimes_{m=1}^N
\bra{a_m}\right)\left(\bigotimes_{m=1}^N
H\right)T^r_N(x,u)\left(\bigotimes_{m=1}^N\ket{k_m}\right)\right]\nonumber\\
& & \times\left[ \left(\bigotimes_{m=1}^N
\bra{b_m}\right)\left(\bigotimes_{m=1}^N
H\right)T^r_N(y,v)R_N(x)\left(\bigotimes_{m=1}^N\ket{k_m}\right)\right]\nonumber\\
& &\otimes\left[\left(\bigotimes_{m=1}^N
\mathfrak{r}(b_m)\right)R_N(y)\left(\bigotimes_{m=1}^N\mathfrak{r}(a_m)\right)R_N(x)
\left(\bigotimes_{m=1}^N\ket{k_m}_{Y_m}\right)\right].\eeqa Now, we
have seen that it is similar to the cases of one qubit as well as
the cases of multiqubits with the controllers, the proof skills have
been shown in the last of Secs. \ref{sec3} and \ref{sec4}. Firstly,
in terms of Eqs.(\ref{tn1}) and (\ref{rn}), we have \beqa
\ket{\psi^{\rm final}_N}&\propto&
\left(\bigotimes_{m=1}^N\ket{a_m}_{A_m}\right)
\otimes\left(\bigotimes_{m=1}^N\ket{c_m}_{C_m}\right)\sum_{j_1\cdots
j_N=0}^1\sum_{k_1\cdots k_N=0}^1\sum_{l_1\cdots
l_N=0}^1 u_{j_1\cdots j_N}y_{k_1\cdots k_n}\nonumber\\
& &\times\left[\left(\bigotimes_{m=1}^N
\bra{a_m}\right)\left(\bigotimes_{m=1}^N
H\right)\left(\bigotimes_{m=1}^N\ket{j_m}\right)\left(\prod_{m=1}^N(-1)^{a_m l_m}\right)\right]\nonumber\\
& &\times\left[
\left(\bigotimes_{m=1}^N\bra{j_m}\right)R_N(x)\left(\bigotimes_{m=1}^N\ket{k_m}\right)\right]\left[
\left(\bigotimes_{m=1}^N\bra{l_m}\right)R_N(x)\left(\bigotimes_{m=1}^N\ket{k_m}\right)\right]
\nonumber\\
& & \times\left[\left(\bigotimes_{m=1}^N
\bra{b_m}\right)\left(\bigotimes_{m=1}^N
H\right)T^r_N(y,v)R_N(x)\left(\bigotimes_{m=1}^N\ket{k_m}\right)\right]\nonumber\\
& &\otimes\left[\left(\bigotimes_{m=1}^N
\mathfrak{r}(b_m)\right)R_N(y)
\left(\bigotimes_{m=1}^N\ket{l_m}_{Y_m}\right)\right].\eeqa
Secondly, from the Eqs.(\ref{rnor}) and (\ref{hrr}), it follows that
\beqa \ket{\psi^{\rm final}_N}&\propto&
\left(\bigotimes_{m=1}^N\ket{a_m}_{A_m}\right)
\otimes\left(\bigotimes_{m=1}^N\ket{c_m}_{C_m}\right)\sum_{j_1\cdots
j_N=0}^1\sum_{k_1\cdots k_N=0}^1u_{j_1\cdots j_N}y_{k_1\cdots k_n}\nonumber\\
& &
\times\left[\left(\bigotimes_{m=1}^N\bra{j_m}\right)R_N(x)\left(\bigotimes_{m=1}^N\ket{k_m}\right)\right]
\left[\left(\bigotimes_{m=1}^N
\bra{b_m}\right)\left(\bigotimes_{m=1}^N
H\right)T^r_N(y,v)R_N(x)\left(\bigotimes_{m=1}^N\ket{k_m}\right)\right]\nonumber\\
& &\otimes\left[\left(\bigotimes_{m=1}^N
\mathfrak{r}(b_m)\right)R_N(y)
\left(\bigotimes_{m=1}^N\ket{j_m}_{Y_m}\right)\right].\eeqa Thirdly,
inserting the complete relation after $T^r_N(y)$ and using Eq.
(\ref{rnor}), we will eliminate a $R_N(x)$ \beqa \ket{\psi^{\rm
final}_N}&\propto& \left(\bigotimes_{m=1}^N\ket{a_m}_{A_m}\right)
\otimes\left(\bigotimes_{m=1}^N\ket{c_m}_{C_m}\right)\sum_{j_1\cdots
j_N=0}^1\sum_{k_1\cdots k_N=0}^1u_{j_1\cdots j_N}y_{k_1\cdots k_n}\nonumber\\
& &
\times\left[\left(\bigotimes_{m=1}^N\bra{j_m}\right)R_N(x)\left(\bigotimes_{m=1}^N\ket{k_m}\right)\right]
\left[\left(\bigotimes_{m=1}^N
\bra{b_m}\right)\left(\bigotimes_{m=1}^N
H\right)T^r_N(y,v)\left(\bigotimes_{m=1}^N\ket{j_m}\right)\right]\nonumber\\
& &\otimes\left[\left(\bigotimes_{m=1}^N
\mathfrak{r}(b_m)\right)R_N(y)
\left(\bigotimes_{m=1}^N\ket{j_m}_{Y_m}\right)\right].\eeqa
Fourthly, substituting Eqs.(\ref{tn1}) and (\ref{rn}) gives \beqa
\ket{\psi^{\rm final}_N}&\propto&
\left(\bigotimes_{m=1}^N\ket{a_m}_{A_m}\right)
\otimes\left(\bigotimes_{m=1}^N\ket{c_m}_{C_m}\right)\sum_{i_1\cdots
i_N=0}^1\sum_{j_1\cdots j_N=0}^1\sum_{k_1\cdots
k_N=0}^1\sum_{l_1\cdots
l_N=0}^1 v_{i_1\cdots i_N}u_{j_1\cdots j_N}y_{k_1\cdots k_n}\nonumber\\
& &\times\left[
\left(\bigotimes_{m=1}^N\bra{j_m}\right)R_N(x)\left(\bigotimes_{m=1}^N\ket{k_m}\right)\right]
\left[\left(\bigotimes_{m=1}^N
\bra{b_m}\right)\left(\bigotimes_{m=1}^N
H\right)\left(\bigotimes_{m=1}^N\ket{i_m}\right)\right]\left(\prod_{m=1}^N(-1)^{b_m
l_m}\right)\nonumber\\
& &\times\left[\left(\bigotimes_{m=1}^N\bra{i_m}\right)
R_N(y)\left(\bigotimes_{m=1}^N\ket{j_m}\right)\right]\left[\left(\bigotimes_{m=1}^N\bra{l_m}\right)
R_N(y)\left(\bigotimes_{m=1}^N\ket{j_m}\right)\right]\otimes
\left(\bigotimes_{m=1}^N\ket{l_m}_{Y_m}\right).\eeqa Finally, from
Eqs.(\ref{rnor}) and (\ref{hrr}), it follows that \beqa
\ket{\psi^{\rm final}_N}&\propto&
\left(\bigotimes_{m=1}^N\ket{a_m}_{A_m}\right)
\otimes\left(\bigotimes_{m=1}^N\ket{c_m}_{C_m}\right)\sum_{i_1\cdots
i_N=0}^1\sum_{j_1\cdots j_N=0}^1\sum_{k_1\cdots
k_N=0}^1 v_{i_1\cdots i_N}u_{j_1\cdots j_N}y_{k_1\cdots k_n}\nonumber\\
& & \times\left[
\left(\bigotimes_{m=1}^N\bra{j_m}\right)R_N(x)\left(\bigotimes_{m=1}^N\ket{k_m}\right)\right]
\left[\left(\bigotimes_{m=1}^N\bra{i_m}\right)
R_N(y)\left(\bigotimes_{m=1}^N\ket{j_m}\right)\right]\otimes
\left(\bigotimes_{m=1}^N\ket{i_m}_{Y_m}\right)\nonumber\\
&=&\left(\bigotimes_{m=1}^N\ket{a_m}_{A_m}\right)
\otimes\left(\bigotimes_{m=1}^N\ket{c_m}_{C_m}\right)\otimes
T^r_N(y,v)T^r_N(x,u)\ket{\xi}_{Y_1\cdots Y_N}, \eeqa where we have
used the fact that \beqa T^r_N(y,v)T^r_N(x,u)\ket{\xi}_{Y_1\cdots
Y_N}&=& \sum_{j_1,\cdots,j_N=0}^1\sum_{k_1,\cdots, k_N=0}^1
u_{j_1\cdots
j_N}y_{k_1\cdots k_N}\nonumber\\
& &
\times\left[\left(\bigotimes_{m=1}^N\bra{j_m}\right)R_N(x)\left(\bigotimes_{m=1}^N\ket{k_m}\right)\right]
T^r_N(y,v)\left(\bigotimes_{m=1}^N\ket{j_m}_{Y_m}\right)\nonumber\\
&=& \sum_{i_1\cdots i_N=0}^1\sum_{j_1\cdots j_N=0}^1\sum_{k_1\cdots
k_N=0}^1v_{i_1\cdots i_N}u_{j_1\cdots j_N}y_{k_1\cdots k_n}\left[
\left(\bigotimes_{m=1}^N\bra{j_m}\right)R_N(x)\left(\bigotimes_{m=1}^N\ket{k_m}\right)\right]\nonumber\\
& & \times\left[\left(\bigotimes_{m=1}^N\bra{i_m}\right)
R_N(y)\left(\bigotimes_{m=1}^N\ket{j_m}\right)\right]\otimes
\left(\bigotimes_{m=1}^N\ket{i_m}_{Y_m}\right). \eeqa Therefore,
after restoring the coefficient, adding the other subspaces and
rearranging the space structure, we finish the proof of our protocol
with two senders in the cases of $N$ qubits.

\end{appendix}

\end{document}